\documentclass[twocolumn]{aastex701} 
\usepackage{color}
\usepackage[titletoc]{appendix}
\usepackage{amsmath}
\usepackage{amssymb}
\usepackage{mathtools}
\usepackage{upgreek}
\usepackage{float}
\usepackage{comment}
\usepackage{enumitem}
\usepackage{natbib}
\usepackage{graphicx}
\usepackage{bm}
\usepackage{totcount}
\usepackage{multirow}
\usepackage{hyperref}
\usepackage{cleveref}
\usepackage{tikz}
\usepackage{soul}
\usepackage{xcolor}
\usepackage{placeins}

\usepackage[most]{tcolorbox}

\newtotcounter{citnum} 
\def\oldbibitem{} \let\oldbibitem=\bibitem
\def\bibitem{\stepcounter{citnum}\oldbibitem}

\shortauthors{Suar et al.}
\shorttitle{Modeling Class 0/I Disks}

\begin{document}

\title{Early Planet Formation in Embedded Disks (eDisk). XXV. Inclination-Induced Minor-Axis Brightness Asymmetries Reveal Limited Dust Settling in Embedded Protostellar Disks}

\author[0009-0000-3694-8993]{Sidhant Kumar Suar}
\affiliation{Department of Astronomy, University of Virginia, Charlottesville, VA 22904, USA}
\affiliation{Virginia Institute of Theoretical Astronomy, University of Virginia, Charlottesville, VA 22904, USA}
\email{qgx5kn@virginia.edu}

\author[0000-0001-7233-4171]{Zhe-Yu Daniel Lin}
\altaffiliation{Jansky Fellow of the National Radio Astronomy Observatory}
\affiliation{National Radio Astronomy Observatory, 520 Edgemont Road, Charlottesville, VA 22903, USA}
\affiliation{Department of Astronomy, University of Virginia, Charlottesville, VA 22904, USA}
\email{dlin@nrao.edu}

\author[0000-0002-7402-6487]{Zhi-Yun Li}
\affiliation{Department of Astronomy, University of Virginia, Charlottesville, VA 22904, USA}
\affiliation{Virginia Institute of Theoretical Astronomy, University of Virginia, Charlottesville, VA 22904, USA}
\email{zl4h@virginia.edu}

\author[0000-0003-0998-5064]{Nagayoshi Ohashi}
\affiliation{Academia Sinica Institute of Astronomy $\&$ Astrophysics, 11F of Astronomy-Mathematics Building, AS/NTU, No.1, Sec. 4, Roosevelt Rd, Taipei 10617, Taiwan}
\email{ohashi@asiaa.sinica.edu.tw}

\author[0000-0003-0845-128X]{Shigehisa Takakuwa}
\affiliation{Department of Physics and Astronomy, Graduate School of Science and Engineering, Kagoshima University, 1-21-35 Korimoto, Kagoshima, Kagoshima 890-0065, Japan}
\affiliation{Academia Sinica Institute of Astronomy $\&$ Astrophysics, 11F of Astronomy-Mathematics Building, AS/NTU, No.1, Sec. 4, Roosevelt Rd, Taipei 10617, Taiwan}
\email{Takakuwa@sci.kagoshima-u.ac.jp}

\author[0000-0002-6195-0152]{John Tobin}
\affiliation{National Radio Astronomy Observatory, 520 Edgemont Road, Charlottesville, VA 22903, USA}
\email{jtobin@nrao.edu}

\author[0000-0002-4540-6587]{Leslie Looney}
\affiliation{Department of Astronomy, University of Illinois, 1002 West Green St, Urbana, IL 61801, USA}
\email{lwl@illinois.edu}

\author[0000-0002-0554-1151]{Mayank Narang}
\affiliation{Academia Sinica Institute of Astronomy $\&$ Astrophysics, 11F of Astronomy-Mathematics Building, AS/NTU, No.1, Sec. 4, Roosevelt Rd, Taipei 10617, Taiwan}
\email{mayank.narang@jpl.nasa.gov}

\author[0009-0004-9279-780X]{Youngwoo Choi}
\affiliation{Department of Physics and Astronomy, Seoul National University, Gwanak-ro 1, Gwanak-gu, Seoul, 08826, Korea}
\email{ychoi@snu.ac.kr}

\author[0000-0003-4022-4132]{Woojin Kwon}
\affiliation{Department of Earth Science Education, Seoul National University, Gwanak-ro 1, Gwanak-gu, Seoul, 08826, Korea}
\affiliation{SNU Astronomy Research Center, Seoul National University, Gwanak-ro 1, Gwanak-gu, Seoul, 08826, Korea}
\affiliation{The Center for Educational Research, Seoul National University, 1 Gwanak-ro, Gwanak-gu, Seoul 08826, Republic of Korea}
\email{wkwon@snu.ac.kr}

\author[0000-0002-9209-8708]{Patrick Sheehan}
\affiliation{National Radio Astronomy Observatory, 520 Edgemont Road, Charlottesville, VA 22903, USA}
\email{psheehan@nrao.edu}

\author[0000-0002-9143-1433]{Ilseung Han}
\affiliation{Institut de Ciències de l’Espai (ICE-CSIC), Campus UAB, Can Magrans S/N, E-08193 Cerdanyola del Vallès, Catalonia, Spain}
\email{ihan@ice.csic.es}


%
%
\begin{abstract}
How and when dust settles in young protostellar disks is a key open question for the dust concentration needed to form planetesimals and, ultimately, planets. However, directly measuring the vertical dust distribution in embedded (Class 0/I) systems remains challenging. We show that brightness asymmetry along the minor axis of highly inclined disks provides a simple, powerful geometric diagnostic of vertical dust structure. Using radiative transfer modeling with RADMC-3D, we generate synthetic continuum images showing that the observed asymmetry arises naturally from disk inclination, optical depth, and dust scale height. We apply this framework to nine Class 0 and I disks from the ALMA Large Program, Early Planet Formation in Embedded Disks (eDisk), using Markov Chain Monte Carlo (MCMC) fitting. Outflow observations independently validate the inferred near- and far-side geometries: all eight sources with useful outflow constraints agree with the orientations predicted by the dust continuum modeling. Our results indicate that most embedded disks show no strong evidence of significant dust settling, with dust scale heights comparable to the gas scale height. Given that the literature indicates Class II disks tend to be well settled, our results reinforce the notion that significant dust settling occurs during the Class I phase, when deeply embedded Class 0 disks transition to their more revealed Class II counterparts. Intriguingly, the timing of dust settling appears to broadly coincide with the development of widespread dust substructures, suggesting that gravitationally driven vertical dust concentration may have triggered substructure formation.
\\[5pt]
\end{abstract}

\section{Introduction}
\label{sec: Introduction}
As a cold molecular cloud core collapses to form a young star, the conservation of angular momentum leads to the formation of a protostellar disk surrounded by an extended envelope \citep[e.g.,][]{Ulrich1976ApJ, Terebey1984ApJ, Mayer2025MNRAS}. The young protostellar disk and envelope are primarily composed of gas, with a small amount of dust. This disk eventually evolves into a protoplanetary disk (PPD), the birthplace of planets. One typically refers to the protostellar disk in the Class 0/I embedded phases of star formation, whereas the PPD is associated with the Class II phase. Planet formation requires submicron-sized dust particles to grow into larger pebbles that eventually accrete to form planetesimals and rocky planetary cores \citep[e.g.,][]{Testi_2014, Birnstiel2024ARA&A}. Over time, these cores can grow and accrete surrounding gas, becoming gas giants. These disks have a relatively short lifespan of a few million years, and dust aggregation has to overcome several growth barriers such as fragmentation and bouncing as the particles collide with each other \citep[e.g.,][]{Guttler2010A&A, Zsom2010A&A, Hasegawa2021ApJ}. Moreover, large dust grains, unlike the gas molecules, act as a pressureless fluid; hence, they undergo Keplerian rotation, while the gas molecules orbit the central star at sub-Keplerian velocities. The slower-orbiting gas produces a headwind on the dust. Aerodynamic drag removes angular momentum from large grains, causing them to drift rapidly inward. The timescale for the radial drift of solid particles toward the star can be much shorter than the timescale for collisional growth, leading to the so-called meter-sized barrier \citep[e.g.,][]{Weidenschilling1977MNRAS, Nakagawa1986Icar, Barriere2005A&A, Hsu2025MNRAS}.

Streaming instability (SI) \citep[e.g.,][]{Youdin2005ApJ, Johansen2007ApJ} is one of the most widely favored mechanisms for overcoming the various growth barriers that hinder planetesimal formation. Although SI can drive the rapid growth of dust particles, it requires the local dust-to-gas mass ratio to be higher than the standard value of $\sim 0.01$ derived from the interstellar medium \citep[ISM; e.g.,][]{Huhn2025A&A}. One way to achieve this is through vertical dust settling, which increases the dust-to-gas ratio in the midplane \citep[e.g.,][]{Hsu2025MNRAS}. The gas scale height is set by the balance between the gas pressure gradient away from the midplane and stellar gravity toward the midplane, giving the gaseous disk a vertical extent. In contrast, the dust scale height is less straightforward than that of the gas. The dust feels the aerodynamic drag force exerted by the gas, causing the dust to sink vertically towards the midplane, creating conditions conducive to SI and ultimately, planet formation \citep[e.g.,][]{Armitage2015arXiv, Birnstiel2024ARA&A, Villenave2025PASP}.

Given the crucial role of dust settling in planet formation, it is essential to constrain the degree of dust settling observationally. In more evolved Class II disks, high-angular-resolution observations have revealed vertically thin dust layers, indicating substantial settling toward the midplane \citep[e.g.,][]{Pinte2016ApJ, Villenave2020A&A, Villenave2022ApJ, Villenave2025A&A}. These results suggest that by the Class II stage, dust evolution has already progressed significantly, creating conditions favorable for processes such as SI and subsequent planetesimal formation. However, a key open question is when significant dust settling begins during disk formation and early evolution. In particular, it remains unclear whether settling is already well underway in the deeply embedded Class 0/I phases or develops primarily at later Class II stages as disks evolve and their surrounding envelopes disperse. Addressing this question is critical for establishing the timeline of planet formation and for understanding how early disk conditions shape the efficiency and location of planetesimal formation.

\begin{figure}[t]
\centering
    \includegraphics[
      width=0.47\textwidth
    ]
    {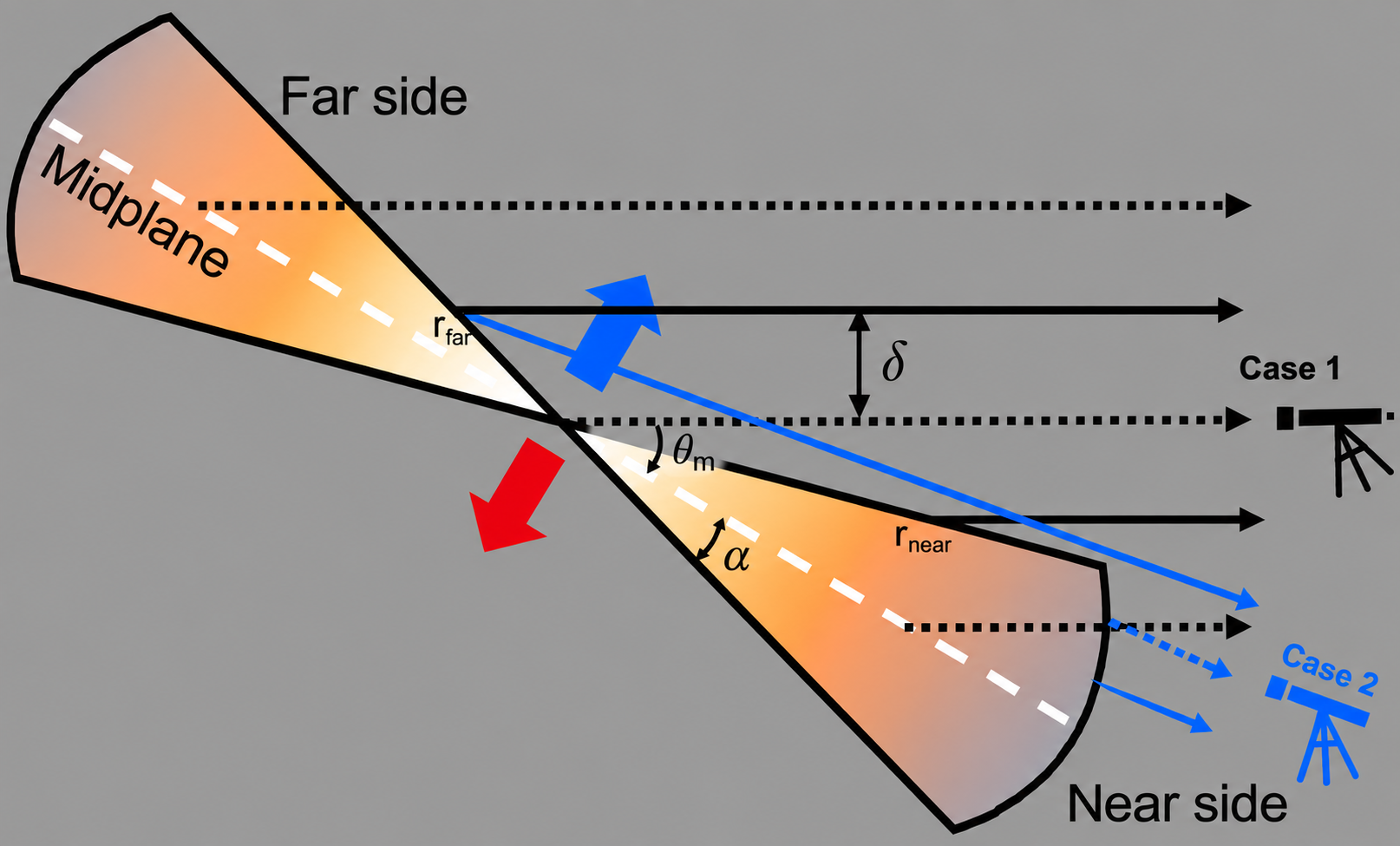}
    \caption{Schematic illustration of the geometric origin of near–far side brightness asymmetry in an inclined dust disk. The white dashed line marks the disk midplane in a meridional cross-section. If the dust is completely settled to the midplane, pairs of sightlines symmetrically displaced about the central line of sight (LOS) intersect the disk at the same radius (e.g., the top and bottom dashed lines), producing a symmetric brightness distribution. For a vertically extended, optically thick dust layer, two distinct cases are illustrated by thick black horizontal lines (Case 1) and thick blue tilted lines (Case 2, more edge-on sightlines). In both cases, the sightlines in the far side are brighter than their symmetric counterparts in the near side (see discussion in the main text). The blue and red arrows indicate the blueshifted and redshifted outflow lobes, respectively; the former is projected onto the far side of the disk, providing an independent observational test of the geometric interpretation of the observed near-far side brightness asymmetry.}
    \label{fig: Disk_Outflow}
\end{figure}

One promising way to probe the vertical thickness of the dust layer in young, embedded disks is to observe the brightness asymmetry along the minor axis of highly inclined systems. This asymmetry has a simple geometric origin. If the dust in an axisymmetric disk has settled completely to the midplane (represented by the white dashed line in the meridional sketch of Figure \ref{fig: Disk_Outflow}), a pair of sightlines symmetrically displaced above and below the central sightline would intersect the midplane at the same distance from the central star (e.g., the top and bottom dashed sightlines in the figure). As a result, the near- and far-sides of the disk would exhibit a symmetric brightness distribution.

The situation changes when the dust layer has a finite vertical thickness and is optically thick, so that only photons originating from the surface layers can reach the observer. There are two distinct cases. In Case 1, pairs of symmetrically displaced sightlines intersect the disk (top or bottom) surface at different radii: the sightline toward the far side (the upper black solid horizontal line) reaches the emitting surface at a smaller radius, denoted by $r_\mathrm{far}$, than the corresponding sightline toward the near side (the lower black solid horizontal line), which reaches the surface at $r_\mathrm{near}$. This relation follows from simple geometry. Let the impact parameter of the symmetric pair of sightlines be $\delta$, i.e., the vertical distance between either solid horizontal sightline and the central dashed sightline through the star in Fig.~\ref{fig: Disk_Outflow}. Denote the inclination angle of the disk midplane relative to the line of sight (LOS) by $\theta_{m} = 90^\circ-\theta_{i}$, where $\theta_{i}$ is the usual disk inclination with respect to the plane of the sky. For an idealized disk with a constant half angular thickness $\alpha$,
\begin{equation}
\label{eq: near_far_geometry}
\begin{split}
    r_\mathrm{near} &= \frac{\delta}{\sin(\theta_{m}-\alpha)}, \qquad
    r_\mathrm{far} = \frac{\delta}{\sin(\theta_{m}+\alpha)}.
\end{split}
\end{equation}
Thus, $r_\mathrm{far} < r_\mathrm{near}$. The difference between the two radii becomes larger as $\theta_{m}$ approaches the disk half angular thickness $\alpha$, corresponding to a more edge-on viewing geometry, and decreases as the disk becomes more face-on (i.e., as $\theta_{m}$ increases). Although the exact relation changes for a flared disk, the qualitative behavior remains the same. Because the disk is generally hotter and brighter at smaller radii, the far side consequently appears brighter than the near side. In Case 2, at a higher disk inclination (i.e., a more edge-on view, illustrated by the thick, tilted blue lines), the far side remains brighter because its emission originates from the warmer top or bottom surface of the disk (the upper solid blue line). In contrast, the near-side emission originates from the cooler radial edge, or side surface, of the disk (the lower solid blue line). In both cases, the asymmetry vanishes if the dust layer is optically thin, since symmetric sightlines would then sample the same dust distribution, differing only in the order in which the material is encountered along the LOS.

A particularly useful feature of this interpretation is that it makes a direct observational prediction. For an inclined disk, the blueshifted outflow lobe (indicated by the blue arrow in Figure \ref{fig: Disk_Outflow}) is projected onto the far side of the disk in the plane of the sky, whereas the redshifted lobe is projected onto the near side. Therefore, if the geometric interpretation is correct, the brighter side of the disk should systematically coincide with the side associated with the blueshifted outflow.

Such minor-axis brightness asymmetries have been observed in several high-resolution ALMA images of embedded disks, including the Class 0 source HH 212 mms \citep[e.g.,][]{Lee2017NatAs} and several sources from the eDisk survey \citep{Ohashi2023ApJ}, such as the Class I disks IRAS 04302$+$2247 \citep{Lin2023ApJ} and R CrA IRS7B \citep{Takakuwa2024ApJ}. Radiative transfer modeling has begun to exploit this geometric effect to constrain the vertical distribution of dust. For example, \cite{Lin2021MNRAS} showed that the Class 0 disk HH 212 mms is consistent with a vertically extended dust layer exhibiting little or no significant settling (see also \citealt{Takakuwa2024ApJ}). Similarly, modeling of the Class I disk IRAS 04302$+$2247 suggests that dust settling, if present, has only recently begun \citep{Lin2023ApJ}. In addition, several other eDisk first-look studies have reported minor-axis brightness asymmetries in Class 0/I disks and have qualitatively discussed their implications for disk vertical structure \citep{Ohashi2023ApJ, Kido2023ApJ, Sai2023, Encalada2024ApJ, Miranda2024A&A, Han2025}. However, no systematic, quantitative modeling of dust vertical structures using minor-axis brightness asymmetries has yet been carried out across a sample of embedded disks. In this paper, we apply the modeling framework developed by \cite{Lin2023ApJ} to a sample of Class 0/I disks from the eDisk survey to place broader constraints on the onset and degree of dust settling during the early stages of disk evolution.

The paper is organized as follows. In Section \ref{sec: RADMC-3D_model}, we present our framework for modeling continuum emission from highly inclined disks and interpret the key features of the synthetic images used to constrain the dust properties, particularly the dust scale height. In Section \ref{sec: ALMA_Modeling}, we model a subset of the eDisk sources, present the fitting results, and discuss the implications of the inferred model parameters. We then discuss the broader implications of our results in Section \ref{sec: Discussion} and summarize our conclusions in Section \ref{sec: Conclusion}.

\section{Dust Continuum Modeling}
\label{sec: RADMC-3D_model}

In this section, we model dust continuum emission from axisymmetric disks using RADMC-3D\footnote{RADMC-3D is publicly available at \url{https://github.com/dullemond/radmc3d-2.0}} \citep{Dullemond2012ascl}. We aim to determine how the brightness asymmetry along the disk's minor axis depends on the disk's inclination $\theta_{i}$, optical depth, and dust scale height. We adopt the modeling framework of \cite{Lin2023ApJ}, which, as mentioned earlier, was successfully applied to one of the eDisk sources, IRAS 04302$+$2247. It is described in the next subsection.

\subsection{Model Setup}
\label{sec: Axi_Disk_Model}

One feature of \cite{Lin2023ApJ}'s modeling framework is that the disk gas density needed for the radiative transfer calculation is parametrized through the disk column density $\Sigma_{g}$, which is, in turn, parametrized through the Toomre $Q$ parameter \citep{Toomre1964ApJ}
\begin{equation}
\label{eq: Toomre}
    Q = \frac{c_{s}\Omega_{\mathrm{K}}}{\pi G\Sigma_{g}},
\end{equation}
where $c_{s} = \sqrt{k_{\mathrm{B}}T/\left(\mu m_{u}\right)}$ is the isothermal sound speed, and $\Omega_{\mathrm{K}} = \sqrt{GM_{\star}/R^{3}}$ the Keplerian frequency. Here, $T$ is the temperature, $\mu \approx 2.29$ is the mean molecular weight, $M_{\star}$ is the stellar mass of the central star, and $R$ is the cylindrical radius. We chose $Q$ to parameterize the Keplerian disk, since gravitational stability requires $Q > 1$.

For a disk in hydrostatic equilibrium, the gas density distribution is
\begin{equation}
\label{eq: rho_g}
\begin{split}
    \rho_{g}(R,z) &= \rho_{g,\mathrm{mid}}\exp{\left(-\frac{z^{2}}{2H_{g}^{2}}\right)},
\end{split}
\end{equation}
where,
\begin{equation}
\label{eq: H_g}
\begin{split}
    H_{g} &= \frac{c_{s}}{\Omega_{\mathrm{K}}}
\end{split}
\end{equation}
is the gas scale height, and $\rho_{g,\mathrm{mid}}$ is the gas density in the midplane.

The gas density in the midplane is related to the column density $\Sigma_{g}$ and gas scale height $H_{g}$, and can be expressed explicitly in terms of the cylindrical radius $R$ as \citep{Lin2021MNRAS}
\begin{equation}
\begin{split}
    \rho_{g,\mathrm{mid}}(R) &= \frac{M_{\star}}{\pi\sqrt{2\pi}Q}R^{-3}.
\end{split}
\end{equation}
Introducing a characteristic radius $R_{0}$, the outer edge of the disk, we can express the midplane mass density as
\begin{equation}
\label{eq: rho_g_mid}
\begin{split}
    \rho_{g,\mathrm{mid}}(R) &= \rho_{g,0}\left(\frac{R}{R_{0}}\right)^{-3},
\end{split}
\end{equation}
where $\rho_{g,0} = M_{\star}/\left(\pi\sqrt{2\pi}R_{0}^{3}Q\right)$ is the characteristic midplane gas density at the outer edge of the disk. We assume $Q$ to be constant over the relatively limited radial range probed by the dust continuum emission.
Following \cite{Lin2023ApJ}, we adopt a power law for the disk temperature,
\begin{equation}
\label{eq: dust_temp}
\begin{split}
    T(R) &= T_{0}\left(\frac{R}{R_{0}}\right)^{-q},
\end{split}
\end{equation}
where $T_{0}$ is the temperature at the outer edge of the disk and the power-law index is fixed at $q = 0.5$. In this initial study, we restrict our modeling to vertically isothermal disks and defer a more self-consistent treatment of the vertical temperature structure to future work. This simplification reduces the number of free parameters and reflects the fact that the vertical temperature distribution remains poorly constrained. In particular, it depends on uncertain grain properties in the disk atmosphere that determine how efficiently stellar radiation is absorbed and reprocessed, as well as on potential accretion heating, which primarily affects the dense inner disk near the midplane (see, e.g., \citealt{Takakuwa2024ApJ}).

As in \cite{Lin2023ApJ}, we assume that the dust scale height $H_{d}(R)$ has the same radial dependence as the gas scale height, leaving the dust scale height at $100~\mathrm{AU}$, $H_{100}$, as a free parameter to be determined by model fitting:
\begin{equation}
\label{eq: H_d}
\begin{split}
    H_{d}(R) &= H_{100}\left(\frac{R}{100~\mathrm{AU}}\right)^{(3-q)/2}.
\end{split}
\end{equation}
With our adopted $q=0.5$, we have $H_d\propto R^{1.25}$, showing that the dust layer is flared, with its thickness at the reference radius of 100 au set by $H_{100}$. For a fixed $H_{100}$, stronger flaring makes the layer thinner at smaller radii and tends to weaken the near–far brightness asymmetry. A larger $H_{100}$ can partly compensate for stronger flaring by restoring a similar thickness where the asymmetry is observed. Thus, the asymmetry primarily constrains the dust thickness in the region it probes, while the fitted $H_{100}$ depends on the adopted flaring index.
A useful comparison is provided by the eDisk source R~CrA~IRS7B-a, which \citet{Takakuwa2024ApJ} modeled independently using a more detailed thermal treatment that includes both stellar irradiation and viscous accretion heating. Their dust disk has an outer radius of $\sim 62~\mathrm{AU}$, close to the $R_0 = 65.9~\mathrm{AU}$ inferred from our modeling. \citet{Takakuwa2024ApJ} assumed that the dust scale height follows the gas hydrostatic scale height but did not explicitly report its value at the outer edge of the dust disk. From their self-consistently calculated midplane temperature distribution (their Figure~9a), we estimate $T \simeq 30~\mathrm{K}$ at $R \simeq 62~\mathrm{AU}$. For their adopted stellar mass of $2.9~\mathrm{M_{\odot}}$, this corresponds to a hydrostatic scale height of $H \simeq 3.2~\mathrm{AU}$. Our best-fit $H_{100} = 6.1~\mathrm{AU}$, together with our adopted $H_{d} \propto R^{1.25}$ scaling, gives $H_{d}(62~\mathrm{AU}) = 3.36~\mathrm{AU}$. Thus, the hydrostatic scale height inferred from their calculated midplane temperature distribution is consistent with our inferred value within its uncertainty, suggesting that the vertically isothermal approximation does not strongly bias the inferred dust scale height for this source.
Nevertheless, neglecting vertical temperature stratification remains a potential source of systematic uncertainty that warrants exploration in future work.

Similarly, we assume that the dust vertical distribution follows the same functional form as that of the gas, but with the gas scale height replaced by the dust scale height $H_{d}$:
\begin{equation}
\label{eq: rho_d}
\begin{split}
    \rho_{d}(R,z) &= \rho_{d,\mathrm{mid}}\exp\left(-\frac{z^{2}}{2H_{d}^{2}}\right).
\end{split}
\end{equation}
To relate the dust $(\rho_{d,\mathrm{mid}})$ and gas $(\rho_{g,\mathrm{mid}})$ densities in the midplane, we assume a constant local dust-to-gas mass density ratio of $\eta = \rho_{d}/\rho_{g}$.

Using Equation \ref{eq: rho_g_mid}, the full functional form of the dust density distribution can be expressed as
\begin{equation}
    \rho_{d}(R,z) \approx \eta\rho_{g,0}\left(\frac{R}{R_{0}}\right)^{-3}\exp\left(-\frac{z^{2}}{2H_{d}^{2}}\right).
\end{equation}

As shown in \cite{Lin2023ApJ} and discussed in detail in Section~\ref{sec: ALMA_Modeling} later, our modeling of the ALMA continuum data yields the outer radius $R_{0}$, the dust scale height $H_{d}$, and a characteristic optical depth
\begin{equation}
\label{eq: tau_nu_0}
\begin{split}
    \tau_{\nu,0} &= \rho_{g,0}R_{0}\kappa_{\nu,g},
\end{split}
\end{equation}
where $\kappa_{\nu,g}$ is the opacity (cross-section per gram of gas) due to the dust in the gas. It is related to the dust opacity $\kappa_{\nu,d}$ (cross-section per gram of dust) through the local dust-to-gas mass ratio $\eta$, i.e., $\kappa_{\nu,g}=\eta \kappa_{\nu,d}$.  

One advantage of \cite{Lin2023ApJ}'s framework is that, once the characteristic optical depth $\tau_{\nu,0}$ and dust disk radius $R_{0}$ are determined from model fitting, it allows us to put a lower limit on the opacity $\kappa_{\nu,g}$ through:
\begin{equation}
\label{eq: kappa_gas}
\begin{split}
    \kappa_{\nu,g} &= \pi\sqrt{2\pi}Q\tau_{\nu,0}\left(\frac{R_{0}^{2}}{M_{\star}}\right),
\end{split}
\end{equation}
where the stellar mass $M_{\star}$ can be determined independently, e.g., through the disk Keplerian rotation, and the Toomre parameter $Q$ must be greater than unity.

\subsection{Interpreting Model Images}

Before modeling the observed Class 0/I protostellar disks, we first aim to develop a conceptual understanding of how minor-axis brightness asymmetry arises in an axisymmetric disk viewed at high inclination using RADMC-3D. As illustrated schematically in Fig.~\ref{fig: Disk_Outflow}, the asymmetry is a combination of geometric and optical depth effects. It depends primarily on the disk's inclination angle, the dust layer's optical thickness, and its geometrical thickness. To study each parameter's effects, we vary one while keeping the other two fixed. 

For illustration, we adopt a fiducial disk radius of $R_{0} = 100 \ \mathrm{AU}$, and a characteristic temperature of $T_{0} = 20 \ \mathrm{K}$ at $R_{0}$. Synthetic images are generated for ALMA Band 6 ($1.3 \ \mathrm{mm}$) used in eDisk \citep{Ohashi2023ApJ}.
We show them with logarithmically spaced contours and an inverse hyperbolic sine (asinh) intensity stretch to highlight both the bright inner-disk emission and the faint extended structure, while retaining a fixed colorbar range for fair comparison across models. Our analysis focuses on the characteristic features of the major- and minor-axis brightness profiles of the modeled disks, following approaches adopted in previous observational and modeling studies \citep[e.g.,][]{Villenave2020A&A, Lin2021MNRAS, Tazaki2025ApJ}.

\subsubsection{Varying the Inclination}
\label{sec: vary_inc}
We follow the convention in which the inclination $\theta_{i}$ increases from $0^{\circ}$ to $90^{\circ}$ as the disk becomes more edge-on. To illustrate the effects of the inclination on the degree of brightness asymmetry along the minor axis of the disk, we define a critical inclination $\theta_{c}$ as the inclination when the observer’s LOS to the central protostar just clears the dust scale height at the outer disk radius $R_{0}$, i.e., 
\begin{equation}
\label{eq: crit_ang}
\begin{split}
    \theta_{c} &= 90^{\circ}-\arctan\left(\frac{H_{d}(R_{0})}{R_{0}}\right).
\end{split}
\end{equation}
At this inclination, the sightlines to the near side of the disk intersect the side surface of the (radial) edge of the disk (facing radially outward) at $R_0$ (see the lower-right arc of Fig.~\ref{fig: Disk_Outflow}). In contrast, those on the far side intersect the disk's top or bottom surface. This difference tends to enhance the near-far side brightness asymmetry.

\begin{figure*}[t]
\centering
    \gridline{
    \includegraphics[
      width=0.5\textwidth
    ]{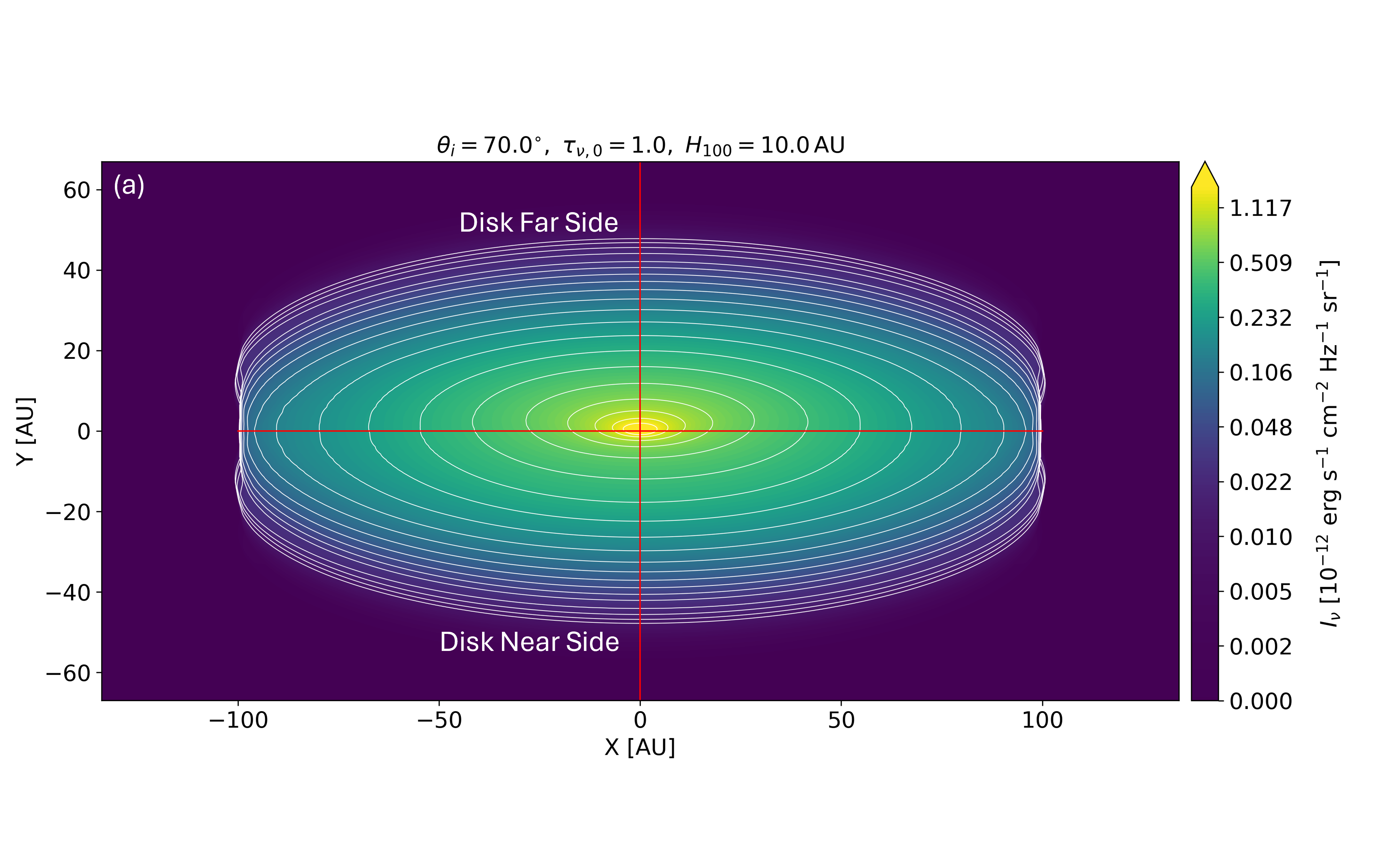}
    \includegraphics[
      width=0.5\textwidth
    ]{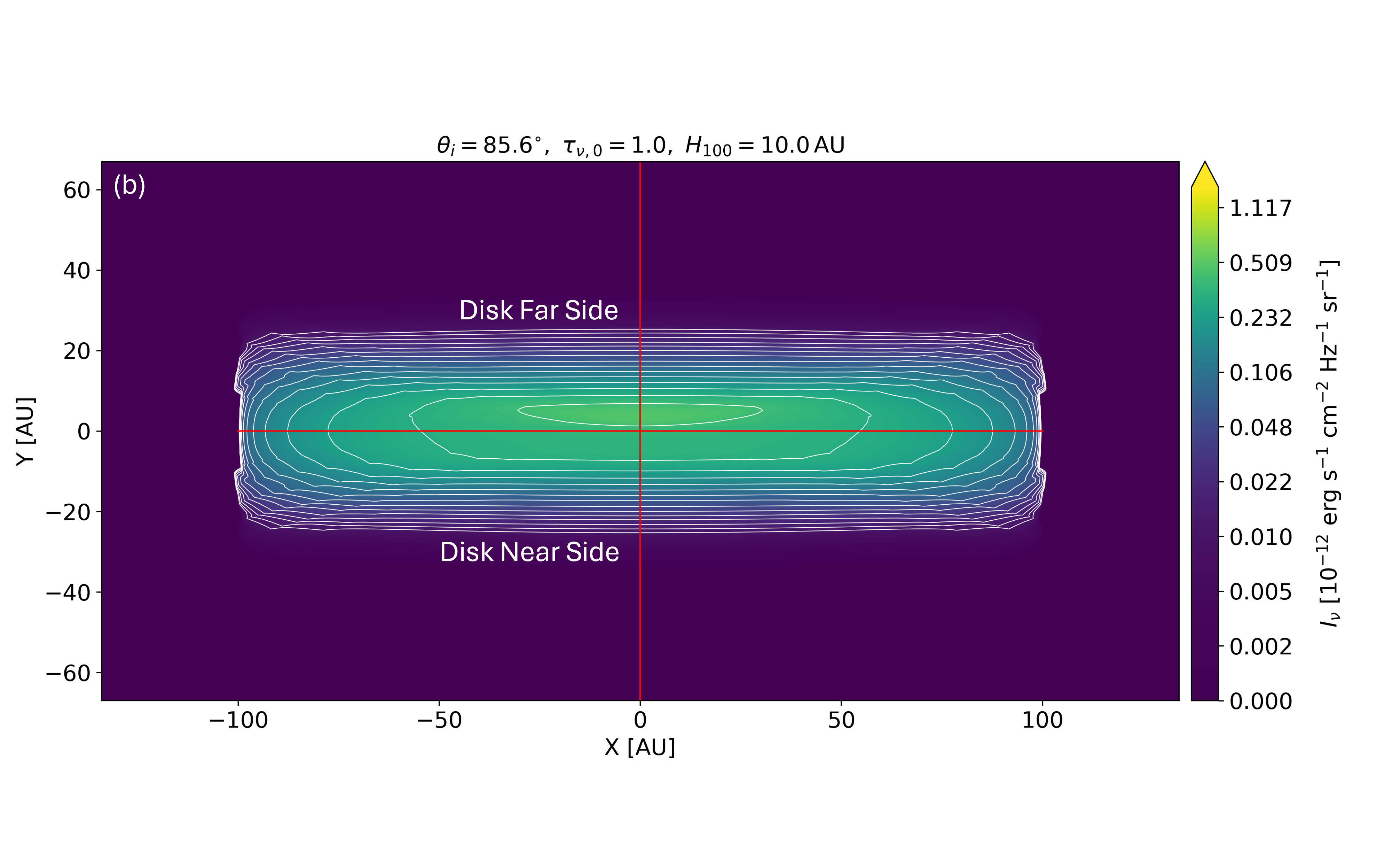}
    }
    \gridline{
    \includegraphics[width=0.5\linewidth]{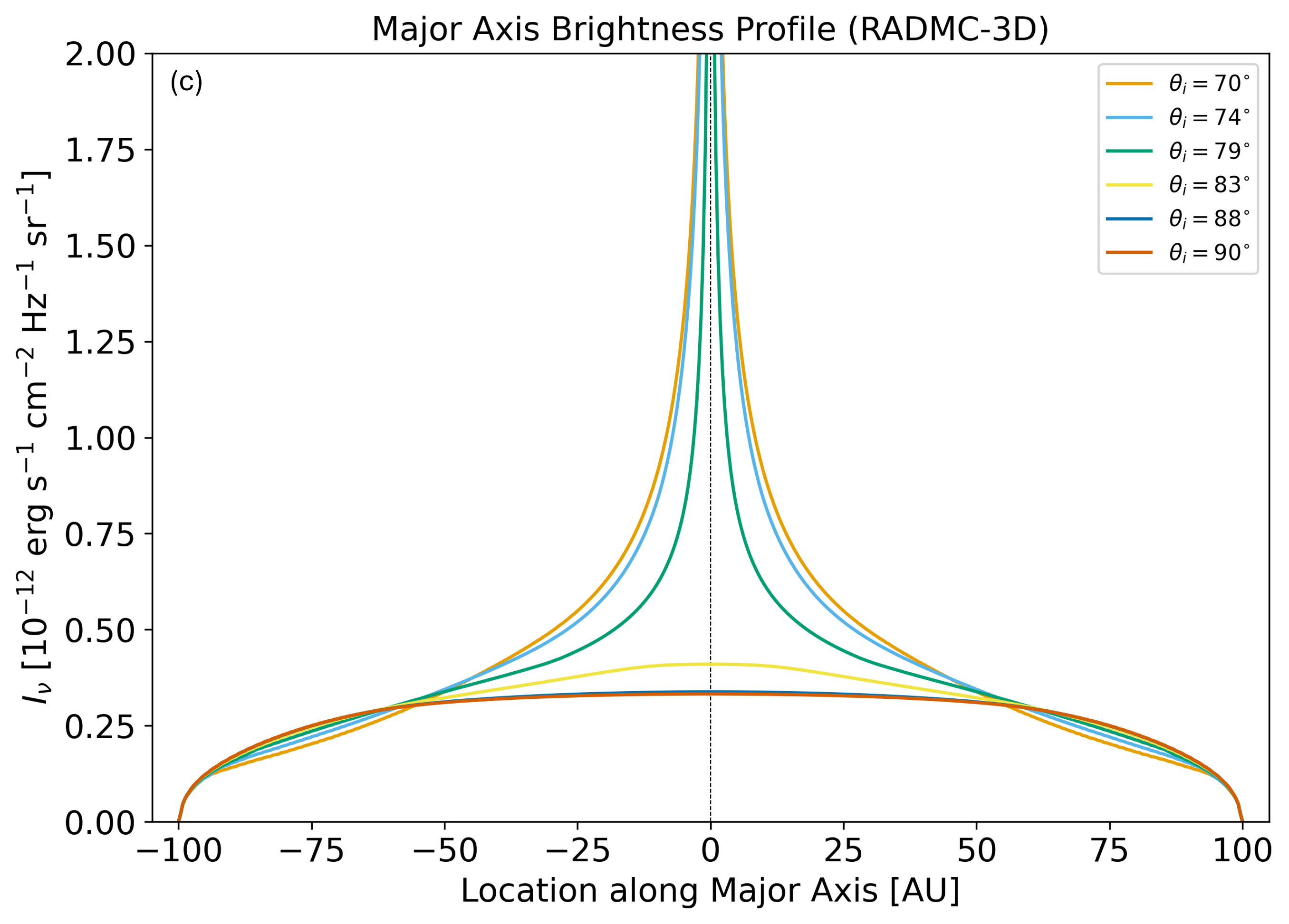}
    \includegraphics[width=0.5\linewidth]{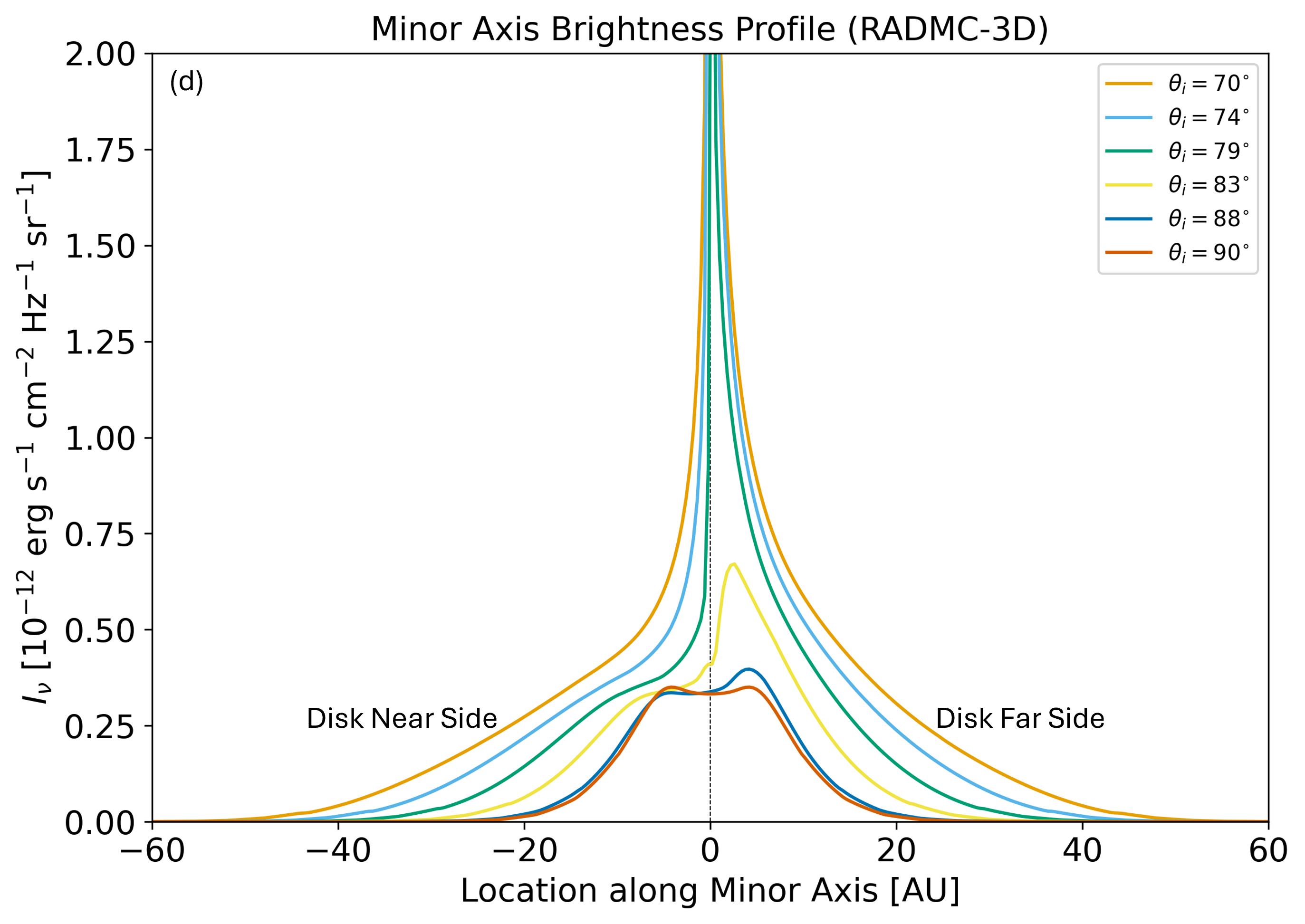}
    }
    \caption{\textit{Top Panels}: Snapshots of the modeled optically thick disk with a fixed reference dust scale height, shown at two different inclination angles: $\theta_{i} = 70^{\circ}$ (a) and $\theta_{i} = 85.6^{\circ}$ (b). Since the distance to the modeled disk is not specified, we present the sky-plane specific intensity $I_{\nu}$, which is independent of distance (also applies to Figures \ref{fig: RADMC3D_tau0} and \ref{fig: RADMC3D_H100}). \textit{Bottom Panels}: Major (c) and minor (d) axis brightness profiles of moderately to highly inclined, optically thick disks with a representative dust scale height.}
    \label{fig: RADMC3D_inc}
\end{figure*}

We vary the inclination angle $\theta_{i}$ from $70^{\circ}-90^{\circ}$, and keep the characteristic optical depth fixed at $\tau_{\nu,0} = 1.0$ and the dust scale height at 100~$\mathrm{AU}$ fixed at $H_{100} = 10 \ \mathrm{AU}$, which corresponds to a critical inclination of $\theta_{c} \approx 84.3^{\circ}$. Figures \ref{fig: RADMC3D_inc} (a) and (b) show the same disk but with two different inclinations: $\theta_{i} = 70^{\circ}$, which is more face-on than $\theta_{c}$, and $85.6^{\circ}$, which is closer to $\theta_{c}$. Clearly, the latter case shows a much more obvious asymmetry along the minor axis than the former. Specifically, the innermost contour of the highest intensity is significantly displaced from the central protostar (marked by the intersection of the two red lines in Fig.~\ref{fig: RADMC3D_inc}b) toward the far side along the minor axis, with a much smaller separation between the innermost intensity contour and the second innermost contour on the far side than on the near side. The brighter sightlines on the far side, compared to their near-side symmetric counterparts, agree with expectations based on simple geometric effects illustrated in Fig.~\ref{fig: Disk_Outflow}.

The brightness asymmetry along the minor axis is further quantified in Figure \ref{fig: RADMC3D_inc}(d) for a set of 6 representative inclination angles. When the disk is more face-on than the critical angle  (with $\theta_{i}$ significantly less than $\theta_{c}$) (as is true for $\theta_{i}=70^\circ$, $74^\circ$, and $79^\circ$), the observer can see the disk surface directly on both the near and far sides, as in Case 1 illustrated in Fig.~\ref{fig: Disk_Outflow}. In this case, the central region is very bright (since the disk does not block it) and the asymmetric brightness profile along the minor axis comes from the fact that the emitting surfaces for the far-side sightlines are located at smaller radii with higher temperatures than their near-side symmetric counterparts (see Fig.~\ref{fig: Disk_Outflow}). Note that the profile becomes more symmetric as the disk becomes more face-on (compare the orange curve for $70^\circ$ and the green curve for $79^\circ$ in Fig.~\ref{fig: RADMC3D_inc}d), 
as expected from the geometric relation derived in Section~\ref{sec: Introduction} (Equation~\ref{eq: near_far_geometry}).

As the inclination angle increases to $\theta_{i}=83^\circ$, the minor-axis brightness profile remains strongly asymmetric (see the yellow curve in Fig.~\ref{fig: RADMC3D_inc}d). Still, for a somewhat different reason from the less inclined cases discussed earlier: emission from the (top) surface of the disk is blocked more by the disk's outer (radial) edge on the near side than on the far side, as illustrated by Case 2 of Fig.~\ref{fig: Disk_Outflow}. Note that the brightness drops more steeply on the far side, which explains why its iso-intensity contours bunch up more closely than on the near side, as noted earlier in Fig.~\ref{fig: RADMC3D_inc}b.  As the disk becomes even more edge-on, the sightlines to the (top or bottom) surfaces on both the near and far sides are blocked by the disk's outer (radial) edge, reducing the asymmetry (see the blue curve for the $88^\circ$ case in Fig.~\ref{fig: RADMC3D_inc}d). In the exact edge-on case of $90^\circ$, the asymmetry disappears completely, as expected.

The brightness profile along the major axis remains symmetric with respect to the center, but its shape changes with the inclination angle (Fig.~\ref{fig: RADMC3D_inc}c). Specifically, the profile becomes boxier as the disk becomes more edge-on, as the central bright region becomes more extincted by the outer parts, making the profile less peaky. Therefore, we expect the disk inclination to be constrained by the brightness distributions along both the minor and major axes. 


\subsubsection{Varying the Optical Depth}
\label{sec: vary_tau0}

\begin{figure*}[t]
\centering
    \gridline{
    \includegraphics[
      width=0.5\textwidth
    ]{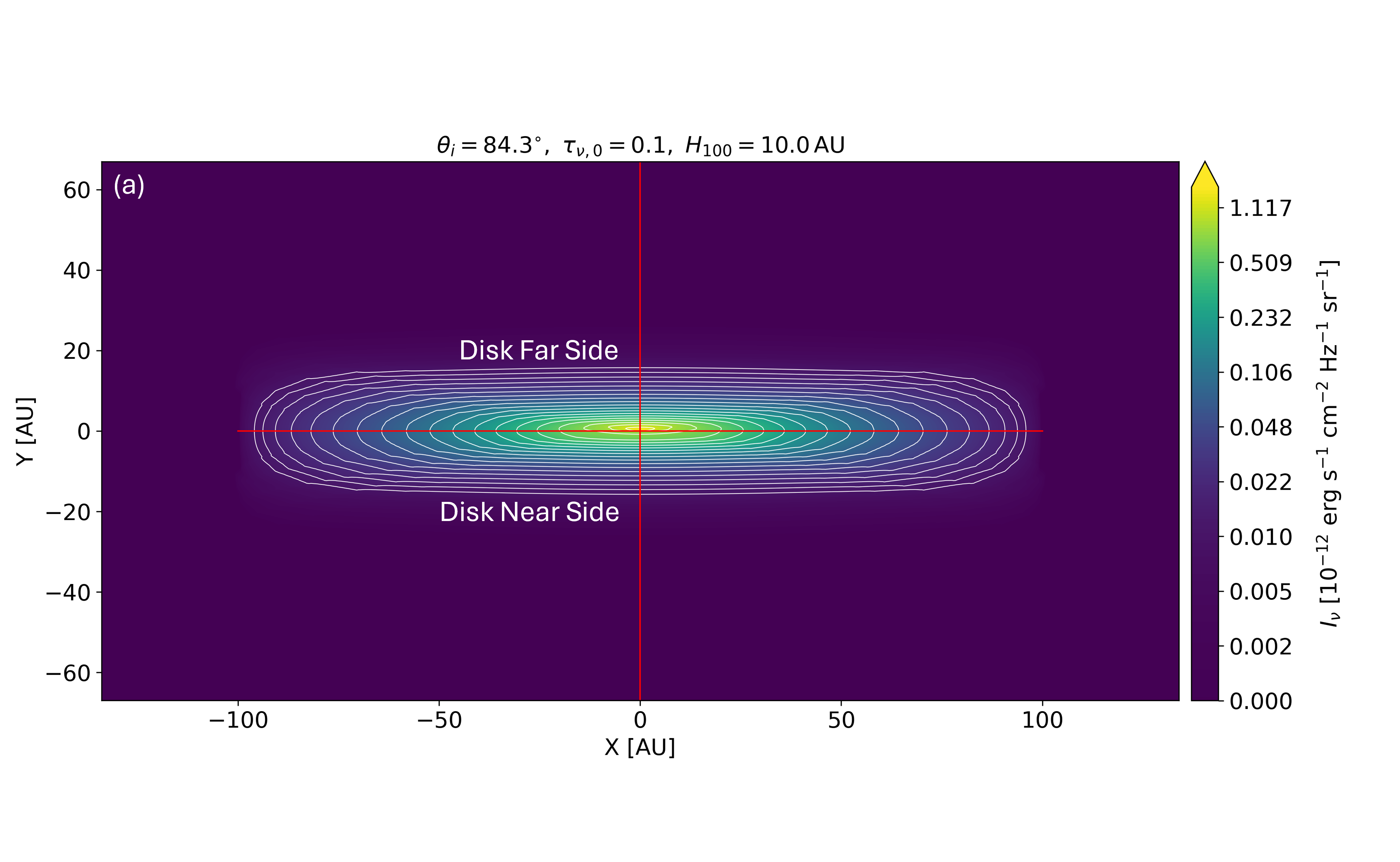}
    \includegraphics[
      width=0.5\textwidth
    ]{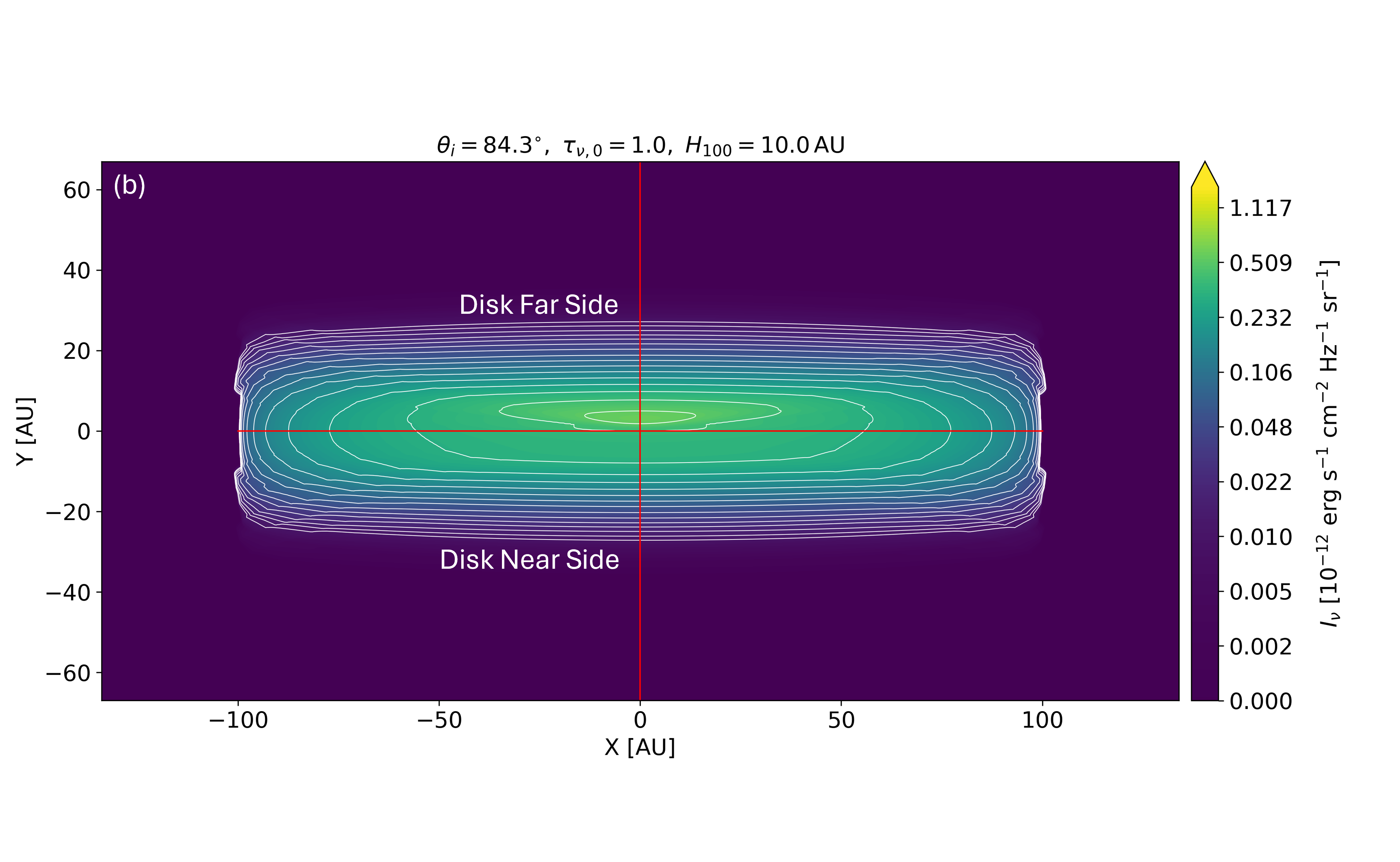}
    }
    \gridline{
    \includegraphics[width=0.5\linewidth]{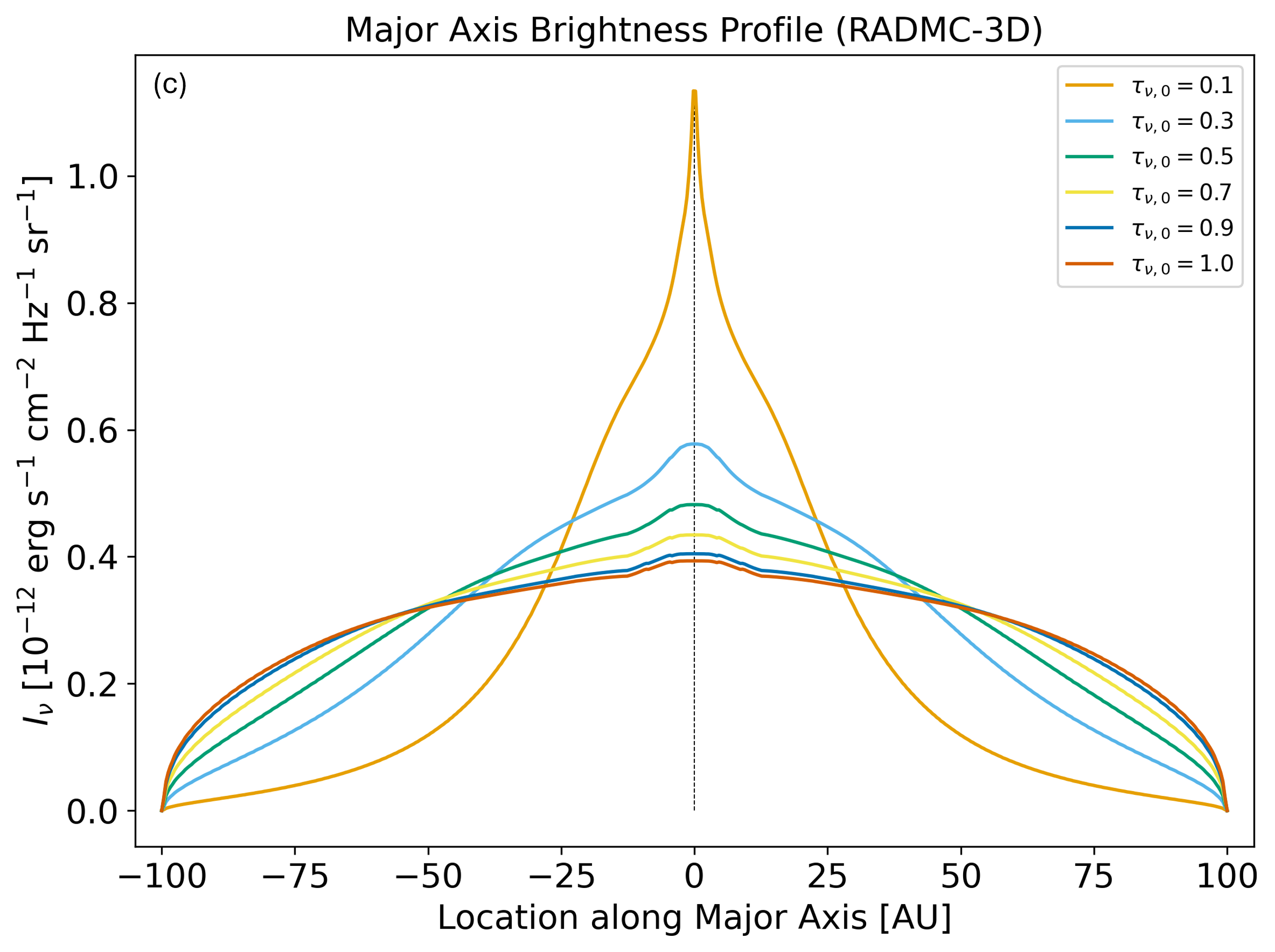}
    \includegraphics[width=0.5\linewidth]{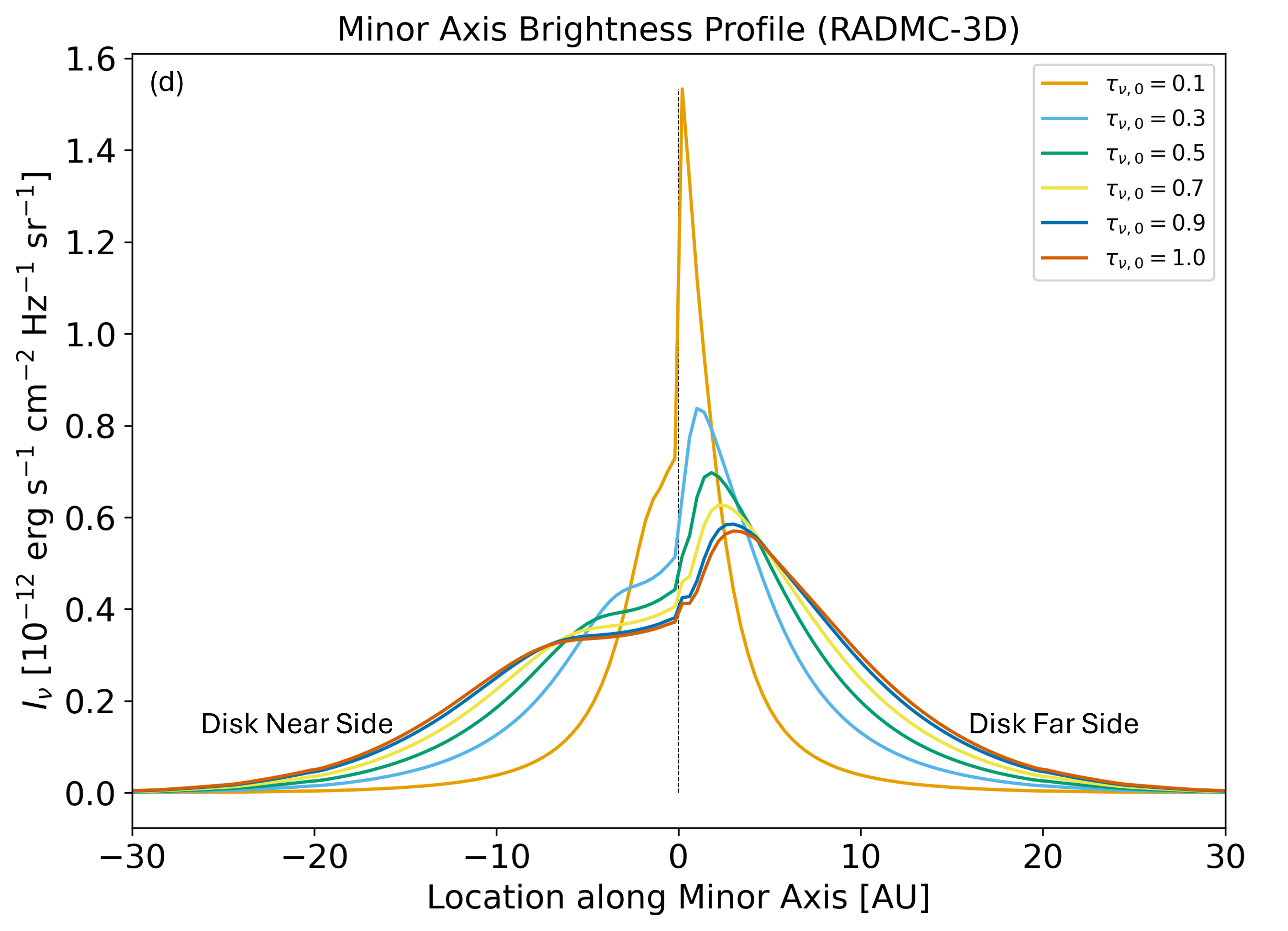}
    }
    \caption{\textit{Top Panels}: Snapshots of the modeled disk, shown at a fixed reference dust scale height and high inclination, for two limiting optical depths corresponding to optically thin (a) and optically thick (b) cases. \textit{Bottom Panels}: Major (c) and minor (d) axis brightness profiles of optically thin to thick, nearly edge-on disks with a representative dust scale height.}
    \label{fig: RADMC3D_tau0}
\end{figure*}

Here, we fix the reference dust scale height at $H_{100} = 10 \ \mathrm{AU}$ and the inclination at the critical angle $\theta_{c} \approx 84.3^{\circ}$, and vary the characteristic optical depth $\tau_{\nu,0}$ from 0.1 to 1.0. It is immediately clear from the intensity maps in the top panels of Fig.~\ref{fig: RADMC3D_tau0} that the brightness asymmetry is much more prominent in the more optically thick case of $\tau_{\nu, 0} = 1.0$, where the brightness peak is offset from the center toward the far side. This offset may be surprising at first sight, since the LOS to the center, where the temperature is highest, clears only one dust scale height at the disk's outer edge, and one might expect the brightness to peak at the center. However, the vertical dust distribution does not stop at a single scale height; it extends beyond it in the disk atmosphere, though it declines rapidly at greater heights. The dust column in the disk atmosphere along the sightline to the central star is high enough to suppress emission from the central region, shifting the brightness peak toward the far side, where temperatures are lower than at the center, but the dust extinction along the LOS is also lower.

The off-center brightness peak is quantified in Figure \ref{fig: RADMC3D_tau0}(d), which shows that, as the disk becomes optically thinner, the brightness peak moves closer to the center (compare, e.g., the red curve for the $\tau_{\nu,0}=1.0$ case with the blue curve for the $\tau_{\nu,0}=0.3$ case). This is not surprising, since one can probe closer to the center through the disk atmosphere as the overall optical depth, characterized by $\tau_{\nu,0}$, decreases. However, even in the most optically thin case plotted, the $\tau_{\nu,0}=0.1$ case, the minor-axis brightness profile remains asymmetric because the sightlines on the near side to the region close to the central star remain optically thick because of high dust densities at small radii, making them dimmer than the far side. Perfect symmetry is achieved only in the limit that all sightlines are optically thin. 

The optical depth has a strong effect on the major-axis brightness profile as well, as shown in Figure \ref{fig: RADMC3D_tau0}c. As the disk becomes optically thicker, the profile transitions from a centrally peaked distribution to a boxier distribution with a flatter top. Notice that the step-function-like behavior of the major axis brightness profile is only evident in the optically thick limit, and this also agrees with the sharp drop observed at the edge of the disk in the major axis profiles in Figures \ref{fig: RADMC3D_inc} and \ref{fig: RADMC3D_H100}. When the disk is optically thinner, there is not enough dust in the outer region to make a substantial contribution to the emission from these regions, and the $\tau_{\nu} = 1$ surface is also much closer to the bright center, making the peak brightness higher. However, as the disk becomes optically thick, the outer regions contribute substantially to the observed emission, as the $\tau_{\nu} = 1$ surface moves outward toward the disk edge, where a sharp drop in intensity is observed. Therefore, the brightness distributions along both minor and major axes can provide constraints on the characteristic optical depth $\tau_{\nu,0}$.


\subsubsection{Varying the Dust Scale Height}
\label{sec: vary_H100}

\begin{figure*}[t]
\centering
    \gridline{
    \includegraphics[
      width=0.5\textwidth
    ]{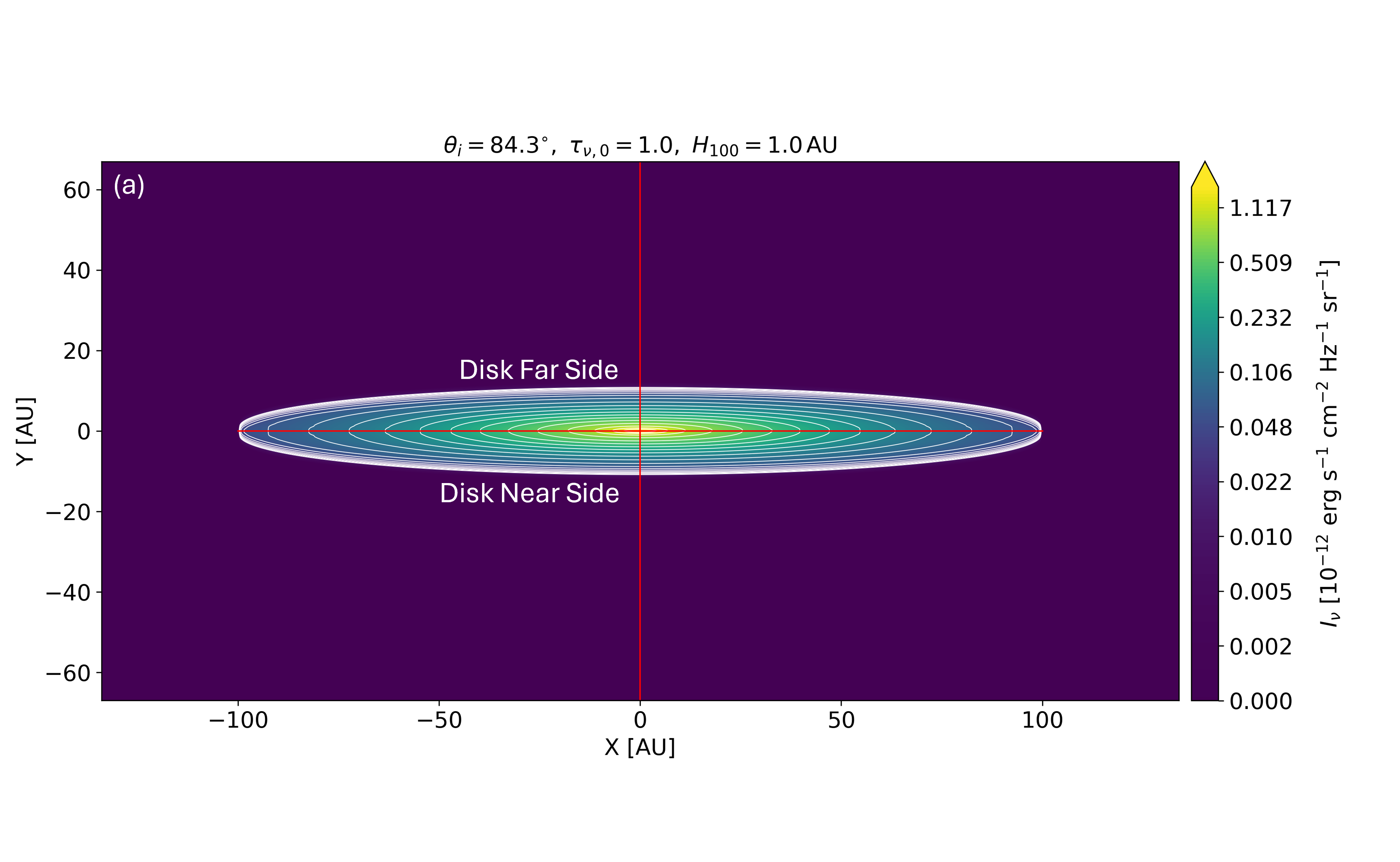}
    \includegraphics[
      width=0.5\textwidth
    ]{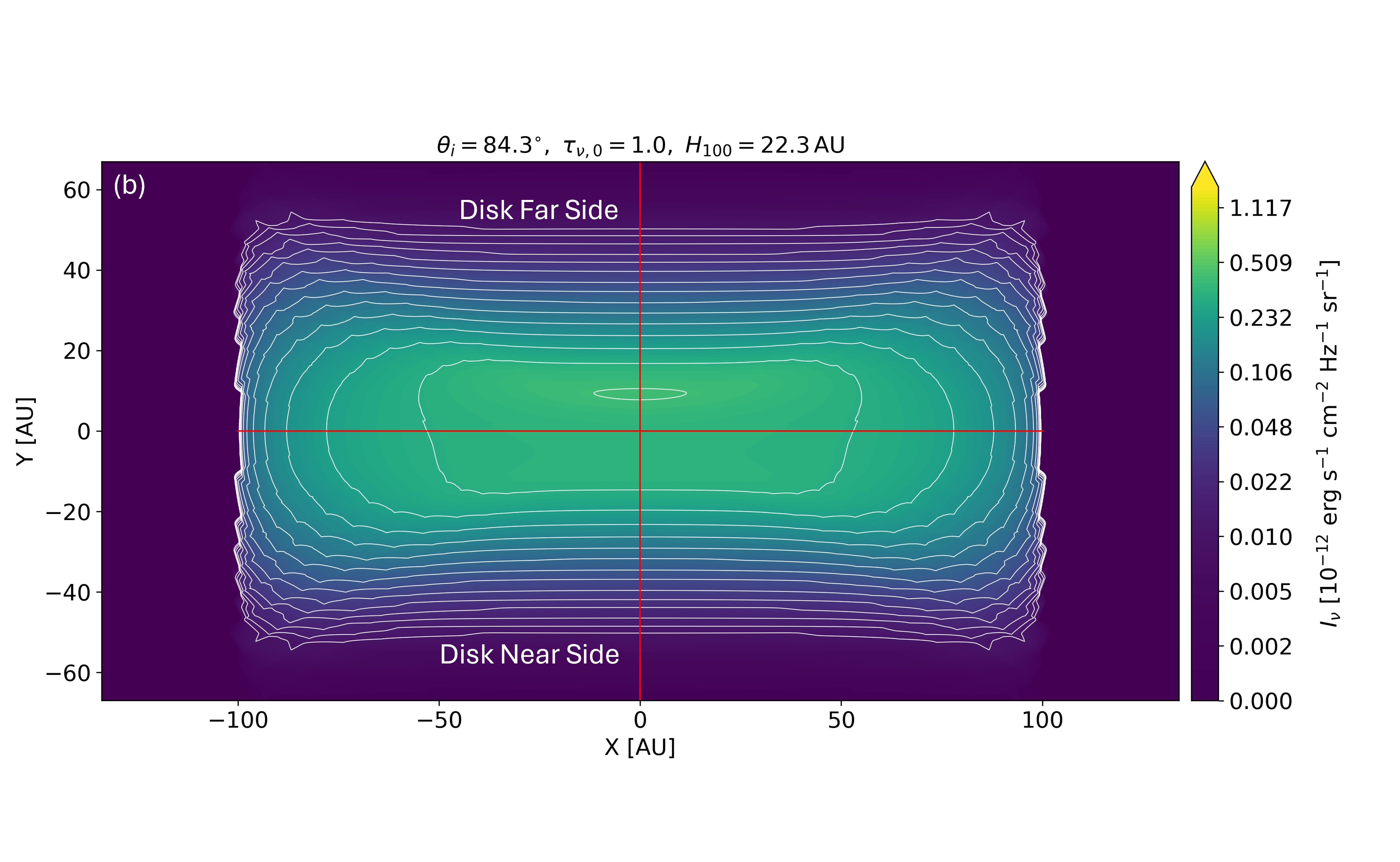}
    }
    \gridline{
    \includegraphics[width=0.5\linewidth]{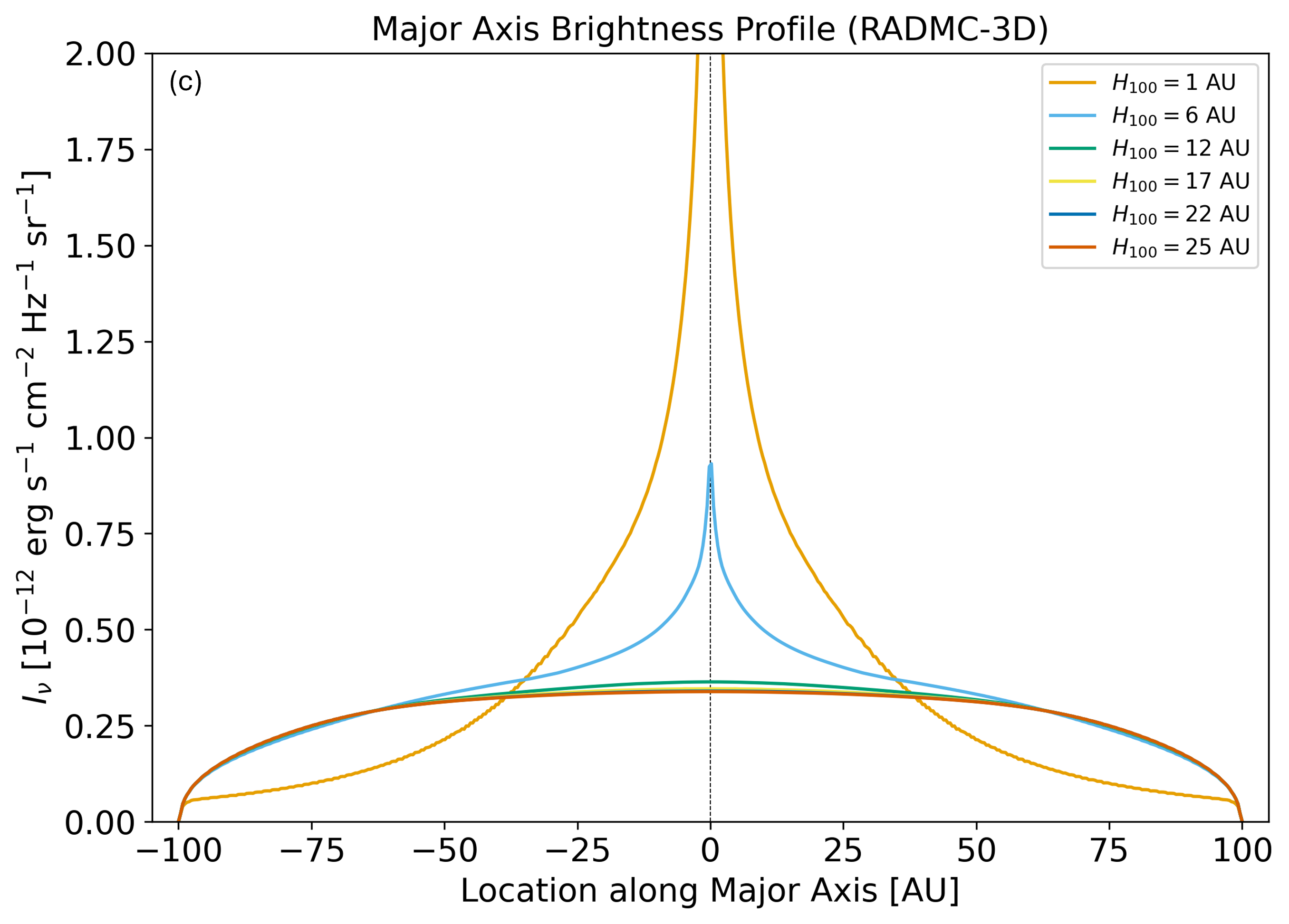}
    \includegraphics[width=0.5\linewidth]{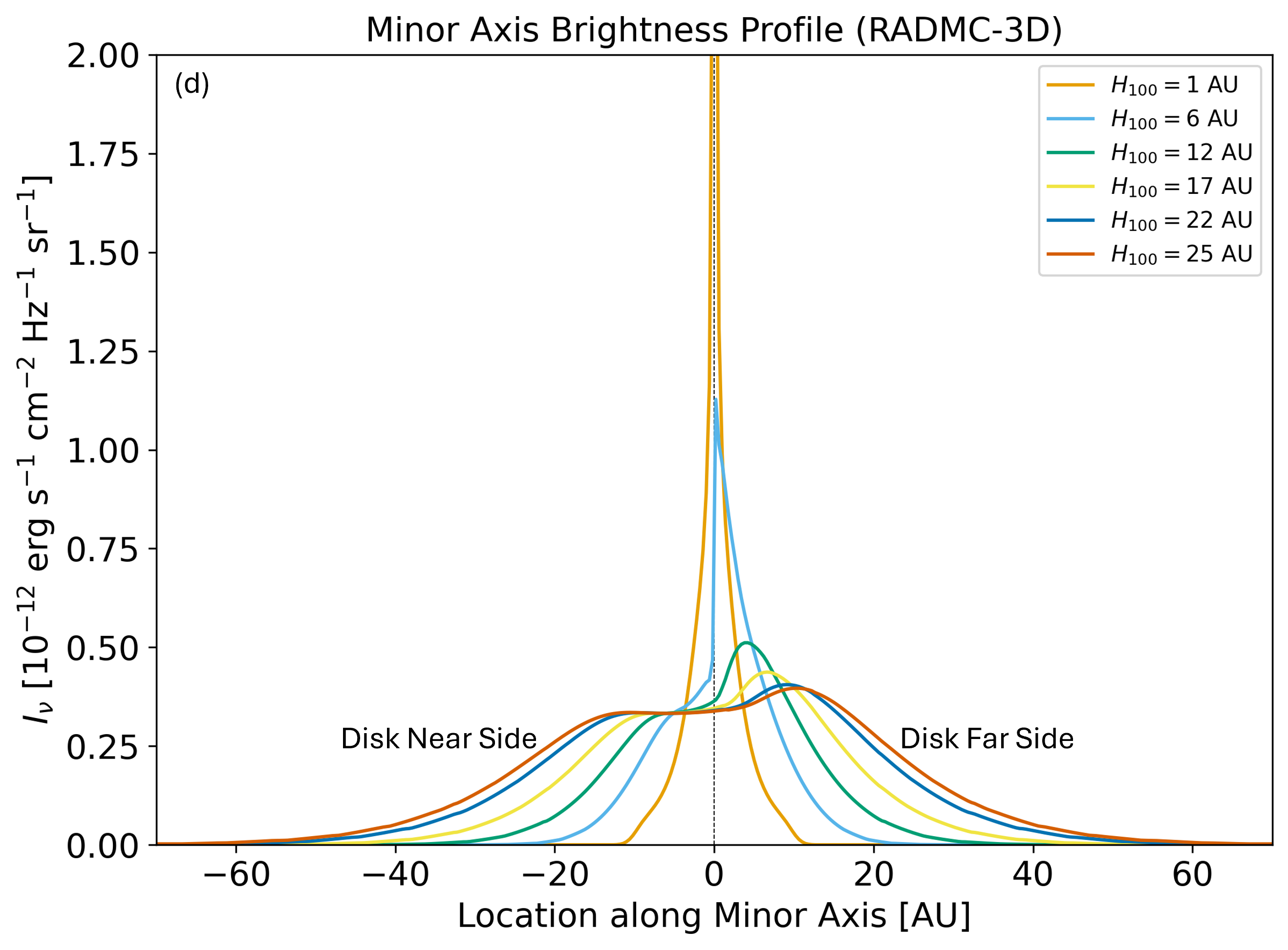}
    }
    \caption{\textit{Top Panels}: Snapshots of the modeled disk, shown in the optically thick and highly inclined regime, for two limiting dust scale heights corresponding to geometrically thin (a) and geometrically thick (b) configurations. \textit{Bottom Panels}: Major (c) and minor (d) axis brightness profiles of geometrically thin to thick, highly inclined, and optically thick disks.}
    \label{fig: RADMC3D_H100}
\end{figure*}
Finally, we vary $H_{100}$, and fix the inclination at $\theta_{c} \approx 84.3^{\circ}$ for a relatively optically thick disk with $\tau_{\nu,0} = 1.0$. The results are presented in Fig.~\ref{fig: RADMC3D_H100}. It is immediately clear from the top-left panel that the intensity distribution in the geometrically thinnest case of $H_{100}=1$~AU is highly symmetric along the minor axis, as expected based on the argument presented in the introduction and illustrated by the white dashed line on the disk midplane in Fig.~\ref{fig: Disk_Outflow}. In contrast, the much thicker case of $H_{100}=22.3$~AU is much more asymmetric, with the brightness peak strongly offset toward the far side. 

Panel (d) shows that the brightness peak shifts toward the center as the dust scale height decreases. For the 4 thickest cases of $H_{100}=12$, $17$, $22$, and $25$~AU, the inclination angle $\theta_{i}$ is less than the critical angle $\theta_c$, with the emission along the sightline to the center suppressed by a large dust optical depth in the disk atmosphere, forcing the brightness to peak on the far side, where sightlines have lower optical depths. As the dust scale height decreases, the optical depth in the disk atmosphere drops, allowing the observer to see the far-side disk surface closer to the center. In the thinner case of $H_{100}=6$~AU, the brightness peak moves close to the center, indicating that the sightlines to the far side become relatively optically thin. In contrast, those to the near side remain optically thick, producing a strong near-far side asymmetry (see the blue curve in panel (d)). The asymmetry is more concentrated near the center than in the thicker-disk cases and may therefore be harder to resolve with a moderate-sized telescope beam.


The dust scale height $H_{100}$ also affects the brightness profile along the major axis, as shown in Figure \ref{fig: RADMC3D_H100}(c). Specifically, it becomes flatter and boxier as $H_{100}$ increases, a trend similar to that from increasing inclination (see Fig.~\ref{fig: RADMC3D_inc}c), and for a similar reason: at a fixed inclination angle, when the dust disk is vertically thin, the observer can see the disk surface on both the near and far sides all the way close to the center, producing highly peaked symmetric brightness profiles along both the major and minor axes, as illustrated by the intensity map for the $H_{100}=1$~AU case (panel a). When the dust layer is vertically thick enough to obscure the disk (top) surface on both the near and far sides, the observer is seeing mostly the emission from the outer (radial) edge (or side surface) of the disk (see the lower solid blue line in Fig.~\ref{fig: Disk_Outflow}), where the temperature is low and approximately constant, yielding a relatively flat (or boxy) profile along the major axis.  

Although inclination, optical depth, and dust scale height influence the observed brightness asymmetry, our results demonstrate that these parameters are not equally effective in producing the observed signal. In particular, while increasing the inclination or optical depth can enhance asymmetry in moderately thick disks, these effects alone are insufficient to produce asymmetry in geometrically thin dust layers. For example, even at high inclinations and large optical depths ($\tau_{\nu,0} \sim 1$), disks with small dust scale heights (e.g., $H_{100} \sim 1 \ \mathrm{AU}$) fail to produce substantial minor-axis near-far-side asymmetry, especially at large enough offsets from the central star that ALMA observations can probe. This indicates that a vertically extended dust distribution is a necessary condition to generate the observed brightness asymmetries. Therefore, while inclination and optical depth help shape the emission morphology, the presence and strength of the asymmetry provide a direct constraint on the dust scale height in highly inclined disks. In the next section, we apply this model to a sample of eDisk sources in \cite{Ohashi2023ApJ} and discuss the results.

\section{Modeling ALMA Observations}
\label{sec: ALMA_Modeling}

\subsection{Sample and fitting method}
\label{subsec: method}

Based on the analysis of the synthetic model images in the previous section, we restrict our target sample to disks with estimated inclinations greater than $65^\circ$, since the brightness asymmetry becomes increasingly difficult to detect, and hence the dust scale height increasingly difficult to constrain, toward lower inclinations (Section~\ref{sec: vary_inc}). We did not impose a single numerical angular resolution threshold. Instead, we assessed the spatial resolution qualitatively on a source-by-source basis based on the synthesized beam size relative to the apparent disk extent, particularly along the minor axis. We selected sources for which the continuum morphology was sufficiently resolved to provide a meaningful constraint on the minor-axis brightness asymmetry, and excluded sources for which beam smearing was likely to prevent a reliable characterization of this asymmetry. We also excluded sources whose irregular dust continuum brightness distributions are difficult to reproduce with the intrinsic axisymmetric disk model described in Section~\ref{sec: Axi_Disk_Model}.
The modeled sources are summarized in Table \ref{tab: vals_sum}, including six Class 0 sources and three Class I sources from \cite{Ohashi2023ApJ}.
To constrain the model parameters, we use an MCMC sampler implemented in the \texttt{emcee} package \citep{Foreman-Mackey2013ascl}. 
To account for the finite angular resolution of the observations, we convolve each synthetic image with a two-dimensional elliptical Gaussian beam whose full widths at half maximum (FWHMs) along the major and minor axes and position angle match those of the synthesized beam for the corresponding observed image. The convolution is performed using the \texttt{fftconvolve} routine from the \texttt{scipy} package. Before comparison, each observed image was subsampled onto a coarser grid with a spacing chosen to be approximately one-half of the synthesized-beam minor-axis FWHM. The corresponding sampling interval was rounded to an integer number of image pixels for each source, thereby preserving uniform grid spacing while reducing computational cost and retaining adequate sampling of the angular resolution. The convolved model image is then interpolated onto the same grid using the \texttt{RectBivariateSpline} interpolation
routine from \texttt{scipy}.
This image-plane approach is particularly useful for spatially resolved binary systems because it separates and fits emission from individual components independently, reducing contamination from nearby companions. To quantify the agreement between the synthetic and observed images at each MCMC iteration, we evaluate the following cost function in the image plane:
\begin{equation}
\label{eq: cost_func}
\begin{split}
    C &= \frac{1}{n}\sum\limits_{i}\left(\frac{o_{i}-m_{i}}{\sigma}\right)^{2},
\end{split}
\end{equation}
where $o_{i}$ and $m_{i}$ are the observed and model intensities at the $i^{\mathrm{th}}$ subsampled pixel, respectively, $\sigma$ is the root-mean-square (rms) noise level of the observation, and $n$ is the number of pixels sampled. The MCMC sampler therefore preferentially explores parameter combinations that produce smaller $C$ values. The reported parameter estimates correspond to the medians of the marginalized posterior distributions.

The intrinsic RADMC-3D image is centered at $\mathrm{\left(X, Y\right)} = \left(0, 0\right)$ in the model coordinate frame, with the positive $\mathrm{Y}$-axis pointing toward the far side of the disk along its minor axis (e.g., see Figure~\ref{fig: RADMC3D_inc}b). To compare the modeled and observed images, we resample the model onto the observed sky grid through the following. For each sky-plane position $\mathrm{\left(X', Y'\right)}$, we compute the corresponding model-frame coordinate $\mathrm{\left(X, Y\right)}$ and interpolate the model image there, according to
\begin{equation}
\label{eq: coord_trans}
\begin{split}
    \mathrm{X} &= \mathrm{\left(X'-X'_{c}\right)}\cos\left(\theta_{\mathrm{rot}}\right)+\mathrm{\left(Y'-Y'_{c}\right)}\sin\left(\theta_{\mathrm{rot}}\right) \\[5pt]
    \mathrm{Y} &= -\mathrm{\left(X'-X'_{c}\right)}\sin\left(\theta_{\mathrm{rot}}\right)+\mathrm{\left(Y'-Y'_{c}\right)}\cos\left(\theta_{\mathrm{rot}}\right),
\end{split}
\end{equation}
where $\mathrm{\left(X', Y'\right)}$ denote the sky-plane coordinates and $\mathrm{\left(X, Y\right)}$ the model-frame coordinates at which the model image is sampled. The quantities ($\mathrm{X'_{c}}$, $\mathrm{Y'_{c}}$) and $\theta_{\mathrm{rot}}$ are free parameters to be constrained by aligning the modeled and observed images in position and orientation, respectively, in the plane of the sky. The modeled and observed images are displayed in the standard sky orientation, in which North corresponds to increasing $\mathrm{Y'}$ and East to decreasing $\mathrm{X'}$. We allow $\theta_{\mathrm{rot}}$ to span $-180^{\circ}$ to $180^{\circ}$, thereby restricting the inclination to $0^{\circ}$--$90^{\circ}$.

Although formal statistical uncertainties can be estimated from the $16^{\mathrm{th}}$ and $84^{\mathrm{th}}$ percentiles of the MCMC posterior distributions, these intervals are typically much smaller than the uncertainties associated with finite angular resolution and image-plane modeling. We therefore estimate the parameter uncertainties independently for each source using one-dimensional cost-function profiles, following the approach of \citet{Lin2023ApJ}. For each fitted parameter, we vary it around the best-fit solution while holding the remaining fitted parameters fixed, and recompute the forward model. Using the cost function $C$ defined in Equation~\ref{eq: cost_func}, we adopt the parameter values at which $C=C_{\min}+1$ on either side of the minimum as the lower and upper uncertainty bounds.

For the derived quantities $H_{d,0}/R_{0}$, $H_{d,0}/H_{g,0}$, and $\kappa_{\nu,g}$, we use standard analytic error propagation. Inspection of the MCMC corner plots shows that, although some correlations are present in the full fitted parameter space, no strong correlations are evident among the parameter combinations that
jointly enter each of these derived quantities. We therefore expect covariance terms to be subdominant and neglect them in the error propagation. The stellar mass $M_{\star}$ is determined independently of the continuum-model parameters. To conservatively account for asymmetric uncertainties, we adopt the larger of the upper and lower uncertainties for each contributing parameter when propagating the errors, yielding symmetric uncertainties for the derived quantities reported in Table~\ref{tab: vals_sum}.

\subsection{Representative example: GSS30 IRS3}
\label{subsec: GSS30}

\begin{figure*}[t]
\centering
    \includegraphics[
      width=1.0\textwidth
    ]{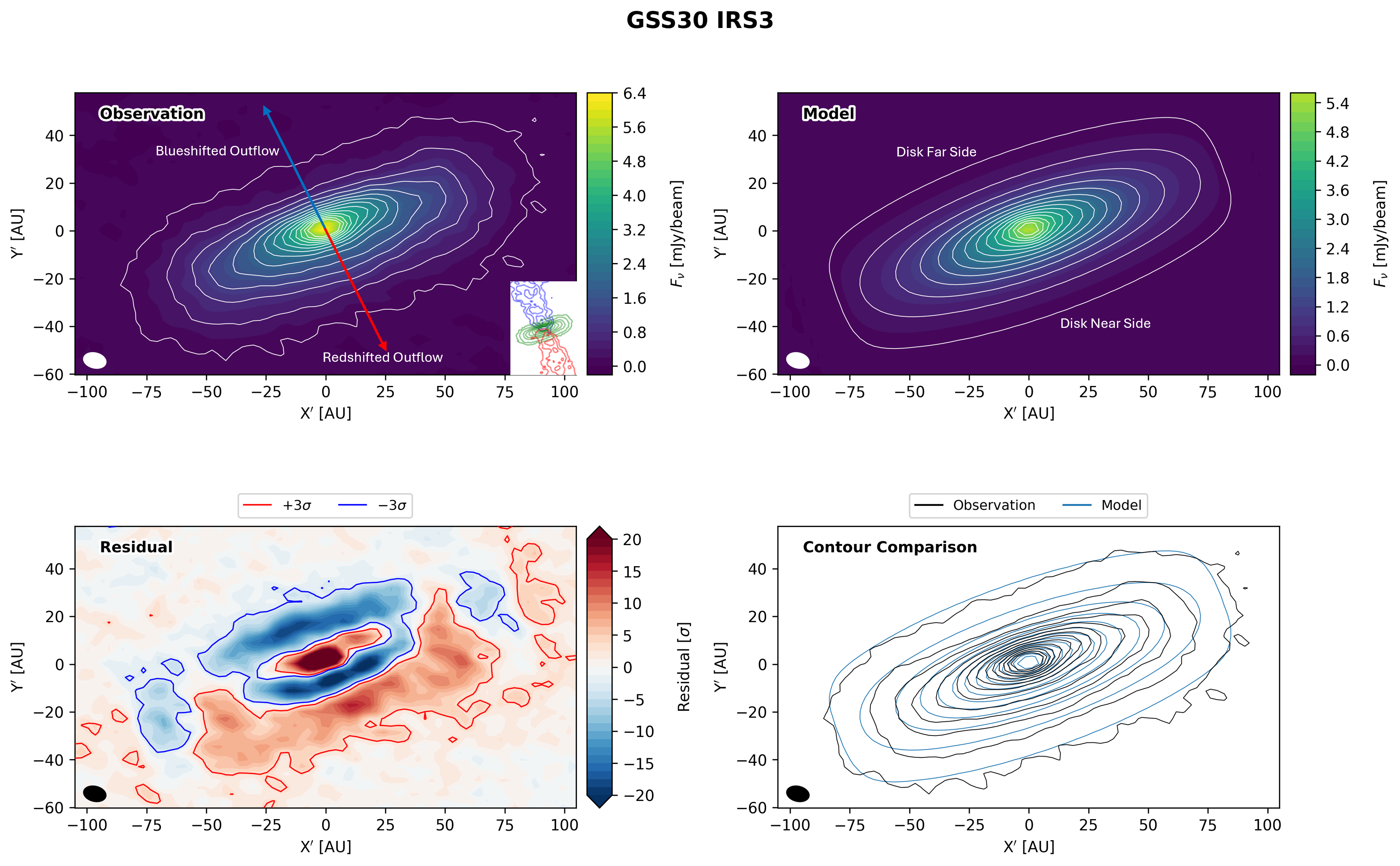}
    \caption{Comparison between the observed and modeled continuum emission for GSS30 IRS3. The \textit{upper-left panel} shows the observed continuum image. The inset shows the CO molecular outflow from \cite{Miranda2024A&A}, with the blue and red arrows indicating the projected directions of the blueshifted and redshifted outflows, respectively. The outflow geometry provides an independent constraint on the disk's near- and far-side orientation. The \textit{upper-right panel} shows the corresponding best-fit RADMC-3D model, with the modeled near and far sides of the disk labeled. The modeled near- and far-side orientations agree with those inferred independently from the observed outflow geometry. The \textit{lower-left panel} displays the residual map, defined as the observed image minus the model image, with red and blue contours marking the $+3\sigma$ and $-3\sigma$ residual levels, respectively. The \textit{lower-right panel} compares the observed (black) and modeled (blue) intensity contours, illustrating the overall agreement in morphology and spatial extent. The same contour levels are used for the observed and modeled images: the lowest contour is set at $5\sigma$, while the remaining contours are linearly spaced up to $80\%$ of the observed peak intensity, corresponding approximately to $[20, 41, \ldots, 230, 250, 271]\sigma$. 
    }
    \label{fig: 2D_RADMC3D_GSS30IRS3}
\end{figure*}


In this section, we present results for a representative case, the Class 0 disk GSS30 IRS3 (Figure \ref{fig: 2D_RADMC3D_GSS30IRS3}); 
the corresponding results for the remaining eight sources are included in Appendix~\ref{sec: Full_Sample}.

The observed continuum emission of GSS30 IRS3 exhibits a clear near-far side asymmetry, with one side appearing brighter and having more closely spaced outer emission contours than the other\footnote{An independent visibility-domain analysis of the same ALMA continuum data by \cite{Narang2026AJ}, using parameterized intensity distributions, identifies a similar near-far side brightness asymmetry as the dominant departure from axisymmetry in GSS30 IRS3.}. The best-fit model successfully reproduces these features, including the offset of the brightness peak along the minor axis and the overall curvature of the emission structure. The modeled near- and far-side orientation is labeled in the upper-right panel of Figure \ref{fig: 2D_RADMC3D_GSS30IRS3}. As an independent check of the modeled disk orientation, we show in the inset in the upper-left panel the highest velocity component of the CO molecular outflow reported by \cite{Miranda2024A&A}, with the corresponding directions of the blueshifted and redshifted outflow lobes indicated by blue and red arrows, respectively, on the continuum map (as also discussed by \citealt{Miranda2024A&A}). The modeled disk orientation agrees with the expectation from the outflow geometry, where the blueshifted outflow is projected onto the far side of the disk in the plane of the sky (see Fig.~\ref{fig: Disk_Outflow}).

The residual map (bottom-left panel) further quantifies the agreement between the model and the observations with 3$\sigma$ contours. The model underpredicts the emission near the center, which could indicate stronger accretion heating at small radii (as explored in detail by \citealt{Takakuwa2024ApJ} for R CrA IRS7B). The residual map shows some spatial coherence, which may indicate substructures too weak to detect with confidence, as stressed by \cite{Narang2026AJ} and Y. Choi et al. (2026, in prep) in their independent continuum modeling of eDisk sources. In any case, the iso-intensity contour comparison in the bottom-right panel shows that the model captures disk morphologies reasonably well, indicating that the adopted axisymmetric structure adequately describes the data.

The continuum images in Figure \ref{fig: 2D_RADMC3D_GSS30IRS3} are displayed using linearly spaced contours to facilitate a direct comparison between the observed and modeled emission. This differs from the synthetic RADMC-3D continuum images presented earlier in Section~\ref{sec: RADMC-3D_model}, for which logarithmically spaced contours and an asinh intensity stretch were used to emphasize both bright inner-disk emission and faint extended structure across models.

As demonstrated by the geometric argument in Section~\ref{sec: Introduction} and confirmed by the radiative transfer models in Section~\ref{sec: vary_H100}, the observed morphology of GSS30 IRS3 cannot be reproduced by a geometrically thin dust disk. While higher inclination and optical depth enhance the near--far side asymmetry, they cannot reproduce the observations without a vertically extended dust layer. The observed asymmetry, therefore, provides a direct constraint on the dust scale height.

For GSS30 IRS3, we infer a disk inclination of $\theta_{i} \approx 73.7^{\circ}$ and a characteristic optical depth of $\tau_{\nu,0} \approx 0.25$, indicating a large but not nearly edge-on inclination and a marginally optically thick disk. The disk is also relatively large, with an outer radius of $R_{0} \approx 86.5 \ \mathrm{AU}$. Most importantly, it has a relatively large dust scale height of $H_{100} = 16.8 \ \mathrm{AU}$ at the reference radius of $R = 100 \ \mathrm{AU}$, corresponding to a dust scale height of $H_{d,0} = 14.0 \ \mathrm{AU}$ at the dust disk outer radius $R_{0} = 86.5 \ \mathrm{AU}$ (based on equation \ref{eq: H_d}), which yields a rather large aspect ratio of $H_{d,0}/R_{0} = 0.16$. It is much larger than, for example, the aspect ratio of the midplane dust layer of $\lesssim 0.01$ inferred in the Class II disk Oph163131 \citep{Villenave2025A&A}. Clearly, the dust probed by ALMA Band 6 observations has yet to settle significantly in this Class 0 disk. For GSS30 IRS3, we find a dust opacity (cross-section per gram of gas including dust) of $\kappa_{\nu,g} \approx 4.6\times10^{-3}Q \ \mathrm{cm^{2}/g}$ at 1.3~mm using equation~(\ref{eq: kappa_gas}), where $Q$ is the Toomre parameter. Since $Q$ is expected to be greater than unity, it provides a lower limit to the opacity.

To quantify the degree of dust settling further, we compare the dust scale height with the gas scale height, which is determined by the stellar mass $M_{\star}$ and the disk temperature $T$. For $M_{\star}$, we adopt the latest dynamical mass inferred from high-resolution line observations that trace the Keplerian disk rotation by Y. Aso et al. (2026, submitted), which is $0.36 \ \mathrm{M_{\odot}}$ for GSS30 IRS3. For temperature, we use the best fit value $T_{0} = 15.5 \ \mathrm{K}$ for the gas temperature at the outer radius $R_{0}$, which yields a gas scale height $H_{g,0} = 10.8 \ \mathrm{AU}$. Intriguingly, the gas and dust scale heights are comparable\footnote{The fact that the dust scale height $H_{d,0}$ is formally somewhat larger than the gas scale height $H_{g,0}$ is likely not significant, given the simplifying model assumptions, such as a vertically isothermal disk.}, again indicating little, if any, dust settling relative to the gas.

\subsection{Results for the full sample}
\label{subsec: Full sample}

The representative example of GSS30 IRS3 demonstrates how fitting the continuum brightness distribution constrains not only the dust scale height and optical depth, but also the near- and far-side orientation of the disk. Applying the same modeling procedure to the remaining eight sources in our sample 
(presented in Appendix~\ref{sec: Full_Sample})
yields a set of best-fit disk parameters for each source, including which side of the disk along the minor axis is the far side. Before discussing the inferred dust properties, it is useful to test whether the modeled disk orientations are consistent with independent observational constraints.

Molecular outflows provide such a test. As discussed in Section \ref{sec: Introduction}, the geometric interpretation of the minor-axis brightness asymmetry predicts that the blueshifted outflow lobe should be projected onto the far side of the disk in the plane of the sky. Outflow information for the modeled sources is summarized in the caption of the figure for each source in 
Appendix~\ref{sec: Full_Sample}
, primarily using measurements reported in the corresponding eDisk source papers and, where available, supplemented by more recent ALMA and JWST observations. Of the nine modeled disks, eight possess sufficiently useful outflow constraints to identify the projected blueshifted and redshifted lobes
\footnote{The outflow orientations for R~CrA~IRS7B-a and R~CrA~IRS7B-b are inferred less directly than for the other six sources; see the corresponding entries in Appendix~\ref{sec: Full_Sample}
for details.}
In all eight cases, the far-side orientation inferred from the dust continuum modeling agrees with that implied by the outflow geometry (see the top panels of the figures in 
Appendix~\ref{sec: Full_Sample}
. The remaining source (R CrA IRS5N) currently lacks sufficiently constraining outflow observations for a meaningful comparison. Future JWST observations may provide an independent test by detecting jets, molecular outflows, or scattered-light reflection nebulae, as recently demonstrated for Ced110 IRS4A. The agreement between the modeled disk orientations and the independently observed outflow geometries strongly supports the geometric interpretation underlying our analysis.

The relatively large dust scale height is not unique to GSS30 IRS3. The best-fit values of $H_{100}$ at the reference radius of $100 \ \mathrm{AU}$ for the $9$ modeled disks range from $5.6$ (for Ced110 IRS4A) to $21.6 \ \mathrm{AU}$ (for R CrA IRAS 32B), with a median value of $\sim 12 \ \mathrm{AU}$ (see the $8^{\mathrm{th}}$ column of Table \ref{tab: vals_sum}). The corresponding aspect ratio of the dust layer at the outer disk edge $H_{d,0}/R_{0}$ ranges from $5\%$ to $16\%$, with a median value of $\sim 9\%$. From Table \ref{tab: vals_sum}, it is clear that the Class 0 disks have a larger median reference dust scale height $H_{100}$ than the Class I disks ($\sim 18$ vs $\sim 6 \ \mathrm{AU}$), with a correspondingly higher median aspect ratio at the outer edge ($\sim 14\%$ vs $\sim 8\%$). Despite the difference, the dust and gas scale height ratio at the outer disk edge $H_{d,0}/H_{g,0}$ is remarkably close to unity for all modeled disks, indicating the dust responsible for the $1.3 \ \mathrm{mm}$ ALMA Band 6 emission has yet to settle significantly for these young disks. 
We emphasize that this conclusion applies specifically to the dust population contributing appreciably to the opacity and emergent continuum emission at ALMA Band 6. A population of substantially larger, centimeter-sized pebbles could be more strongly settled toward the disk midplane while contributing relatively little to the observed 1.3~mm emission, and its vertical distribution would therefore remain poorly constrained by the present analysis. Thus, our results do not rule out significant settling of all solid material; rather, they indicate that the dust population traced by the Band 6 continuum remains vertically extended. If larger grains contribute non-negligibly to the Band 6 opacity, the inferred scale height should instead be interpreted as an effective, opacity-weighted vertical distribution of the emitting grain population.

One reason for the larger dust and gas scale heights in the little-settled Class 0 disks is that they tend to have lower stellar masses (Y. Aso et al. 2026, submitted), which leads to less vertical gravitational compression of the disk gas (and dust) than in the modeled Class I sources. 
Another reason is that these disks often have relatively high temperatures near their outer edges because the surrounding massive envelope absorbs and reprocesses stellar radiation, reradiating part of the energy toward the disk, and thereby providing additional heating to the outer disk (see \citealt{Lin2021MNRAS} for HH 212 mms and \citealt{Hoff2018A&A} for L1527 IRS).

\begin{figure}[t]
\centering
    \includegraphics[
      width=0.5\textwidth
    ]{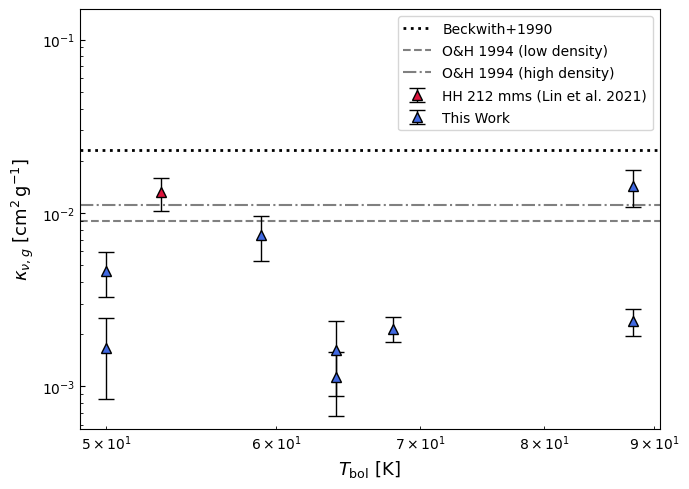}
    \caption{Lower limits on the opacity at 1.3~mm, $\kappa_{\nu,g}$, derived assuming $Q=1$, as a function of bolometric temperature, $T_{\mathrm{bol}}$. Blue triangles denote sources modeled in this work, while the red triangle represents the Class 0 disk HH 212 mms from \cite{Lin2021MNRAS}. Upward-pointing triangles indicate lower limits. The horizontal lines mark commonly adopted opacity values from \cite{Beckwith1990AJ} (dotted line) and \cite{Ossenkopf1994A&A} (dashed and dash-dotted lines for the low- and high-density models, respectively). The dust opacity values reported by \cite{Ossenkopf1994A&A} have been converted to $\kappa_{\nu,g}$ (cross-section per gram of gas including dust); details of this conversion are provided in the text.}
    \label{fig: kappa_g_Disks}
\end{figure}

As explained in Section \ref{sec: Axi_Disk_Model}, one of the advantages of the modeling framework adopted by \cite{Lin2023ApJ} is that we can set a lower limit to the opacity $\kappa_{\nu,g}$ using equation (\ref{eq: kappa_gas}) once the characteristic optical depth $\tau_{\nu,0}$ and the disk outer edge radius $R_{0}$ are determined by model fitting. 
The inferred opacity $\kappa_{\nu,g}$ at 1.3~mm for GSS30 IRS3 is near the middle of the values for the modeled disks, which range from $1.13\times10^{-3}Q$ (for R CrA IRAS 32A) to $14.0\times10^{-3}Q \ \mathrm{cm^{2}/g}$ (for IRAS 04302$+$2247); see the last column of Table \ref{tab: vals_sum} and Figure \ref{fig: kappa_g_Disks}. 
The variation is primarily driven by the factor $M_{\star}/R_{0}^{2}$, which sets the characteristic density-column scale $\rho_{g,0}R_{0} \propto M_{\star}/(Q R_{0}^{2})$. For example, the Class I source IRAS 04302$+$2247 has by far the largest disk ($R_{0} \sim 325~\mathrm{AU}$) among the modeled disks, yielding a much smaller value of $M_{\star}/R_{0}^{2}$ than the other sources. Although its inferred characteristic optical depth $\tau_{\nu,0}$ is among the lowest ($0.29$), a relatively large opacity is still required because $\kappa_{\nu,g} \propto \tau_{\nu,0}R_{0}^{2}/M_{\star}$. Conversely, the Class 0 disk R~CrA~IRAS~32A is relatively compact ($R_{0} \sim 28~\mathrm{AU}$) and has a relatively large stellar mass, yielding a larger value of $M_{\star}/R_{0}^{2}$ and therefore requiring a smaller opacity to produce a characteristic optical depth of order unity.

For reference, in Figure \ref{fig: kappa_g_Disks} we have plotted three opacity values (cross-section per gram of gas including dust) often used in the literature at 1.3~mm: $\kappa_{\nu,g} = 2.3\times10^{-2} \ \mathrm{cm^{2}/g}$ \citep{Beckwith1990AJ}, $8.99\times10^{-3} \ \mathrm{cm^{2}/g}$ (\cite{Ossenkopf1994A&A}; low density) and $1.11\times10^{-2} \ \mathrm{cm^{2}/g}$ (\cite{Ossenkopf1994A&A}; high density), as well as the value inferred for the nearly edge-on Class 0 disk HH 212 mms \citep{Lin2021MNRAS}. The opacity values obtained from \cite{Ossenkopf1994A&A} are originally in terms of the cross-section per gram of dust, $\kappa_{\nu,d}$, but we convert them to $\kappa_{\nu,g}$ assuming a standard dust-to-gas mass ratio of $\eta = 0.01$. 
As discussed in \cite{Lin2023ApJ}, the relatively high lower limits on the opacities (assuming $Q = 1$) for HH 212 mms and IRAS 04302$+$2247 favor the opacity proposed by \cite{Beckwith1990AJ}, as they already exceed the other two opacity estimates. The remaining 6 Class 0 disks and 1 Class I disk (no estimate for R CrA IRS7B$-$b) have lower limits below all three reference values, which is reassuring but does not strongly differentiate among them. Indeed, the majority of these disks have lower limits $\kappa_{\nu,g} \sim 2\times10^{-3} \ \mathrm{cm^{2}/g}$ (assuming $Q = 1$) that are nearly an order of magnitude lower than \cite{Beckwith1990AJ}'s value ($2.3\times10^{-2} \ \mathrm{cm^{2}/g}$). If these early disks are close to gravitational instability (GI), as advocated by \cite{Xu2021MNRAS_I, Xu2021MNRAS_II}, through non-ideal magnetohydrodynamics (MHD) disk formation simulations, the opacity $\kappa_{\nu,g}$ would be an order of magnitude lower than \cite{Beckwith1990AJ}'s value for these disks. Alternatively, their opacities can match \cite{Beckwith1990AJ}'s value (and the other two reference values) if these disks are strongly gravitationally stable, with $Q$ much larger than unity. This is consistent with the lack of obvious large-scale spirals in the continuum image, although a large optical depth may hinder their detection, complicating interpretation.

\begin{deluxetable*}{lcccccccccc}
\scriptsize
\setlength{\tabcolsep}{3pt}
\tablecaption{Parameters from the radiative transfer modeling of selected Class 0/I disks from the eDisk sample.}
\label{tab: vals_sum}
\tablehead{
\colhead{Source} &
\colhead{Class} &
\colhead{$M_{\star} \tablenotemark{a}$} &
\colhead{$R_{0}$} &
\colhead{$T_{0}$} &
\colhead{$\theta_{i}$} &
\colhead{$\tau_{\nu,0}$} &
\colhead{$H_{100}$} &
\colhead{$H_{d,0}/R_{0} \tablenotemark{b}$} &
\colhead{$H_{d,0}/H_{g,0} \tablenotemark{c}$} &
\colhead{$\kappa_{\nu,g}$} \\
\colhead{} &
\colhead{} &
\colhead{$[\mathrm{M_\odot}]$} &
\colhead{$[\mathrm{AU}]$} &
\colhead{$[\mathrm{K}]$} &
\colhead{$[^{\circ}]$} &
\colhead{} &
\colhead{$[\mathrm{AU}]$} &
\colhead{} &
\colhead{} &
\colhead{$[10^{-3}\ \mathrm{cm^{2}/g}]$}
}
\startdata
IRAS 16544$-$1604 & 0 & $0.19_{-0.05}^{+0.09}$ & $26.2_{-0.7}^{+0.6}$ & $54.9_{-1.8}^{+1.9}$ & $78.3_{-1.0}^{+1.0}$ & $0.52_{-0.05}^{+0.06}$ & $21.5_{-1.6}^{+1.7}$ & $0.15_{-0.01}^{+0.01}$ & $0.87_{-0.22}^{+0.22}$ & $1.7_{-0.8}^{+0.8}Q$ \\[5pt]
GSS30 IRS3      & 0 & $0.36_{-0.06}^{+0.10}$ & $86.5_{-1.6}^{+1.1}$ & $15.5_{-0.3}^{+0.3}$ & $73.7_{-0.7}^{+0.7}$ & $0.25_{-0.02}^{+0.02}$ & $16.8_{-0.9}^{+0.9}$ & $0.16_{-0.01}^{+0.01}$ & $1.3_{-0.2}^{+0.2}$ & $4.6_{-1.3}^{+1.3}Q$ \\[5pt]
R CrA IRAS 32A  & 0 & $0.60_{-0.14}^{+0.22}$ & $27.8_{-0.7}^{+0.6}$ & $49.1_{-1.8}^{+1.8}$ & $69.0_{-1.6}^{+1.5}$ & $0.99_{-0.13}^{+0.15}$ & $18.3_{-2.0}^{+2.1}$ & $0.13_{-0.02}^{+0.02}$ & $1.4_{-0.3}^{+0.3}$ & $1.13_{-0.45}^{+0.45}Q$ \\[5pt]
R CrA IRAS 32B  & 0 & $0.29_{-0.07}^{+0.12}$ & $21.2_{-0.6}^{+0.6}$ & $46.9_{-2.0}^{+1.9}$ & $73.7_{-1.4}^{+1.5}$ & $1.19_{-0.19}^{+0.23}$ & $21.6_{-2.3}^{+2.2}$ & $0.15_{-0.02}^{+0.02}$ & $1.24_{-0.29}^{+0.29}$ & $1.6_{-0.7}^{+0.7}Q$ \\[5pt]
Ced110 IRS4A    & 0 & $1.61_{-0.24}^{+0.22}$ & $89.4_{-1.1}^{+0.8}$ & $20.6_{-0.3}^{+0.4}$ & $73.6_{-0.7}^{+0.6}$ & $0.49_{-0.03}^{+0.03}$ & $5.6_{-0.3}^{+0.3}$  & $0.050_{-0.003}^{+0.003}$ & $0.79_{-0.07}^{+0.07}$ & $2.15_{-0.35}^{+0.35}Q$ \\[5pt]
R CrA IRS5N    & 0 & $0.36_{-0.05}^{+0.10}$ & $56.3_{-1.1}^{+0.5}$ & $18.6_{-0.4}^{+0.4}$ & $66.0_{-1.0}^{+1.0}$ & $0.96_{-0.07}^{+0.08}$ & $9.6_{-0.7}^{+0.8}$  & $0.08_{-0.01}^{+0.01}$ & $0.76_{-0.12}^{+0.12}$ & $7.5_{-2.2}^{+2.2}Q$ \\[5pt]
R CrA IRS7B$-$a   & I & $2.71_{-0.26}^{+0.44}$ & $65.9_{-0.5}^{+0.1}$ & $42.8_{-0.7}^{+0.7}$ & $68.0_{-0.6}^{+0.6}$ & $1.7_{-0.1}^{+0.1}$ & $6.1_{-0.3}^{+0.5}$  & $0.050_{-0.005}^{+0.005}$ & $0.84_{-0.10}^{+0.10}$ & $2.4_{-0.4}^{+0.4}Q$ \\[5pt]
R CrA IRS7B$-$b   & I &         ---            & $24.6_{-0.4}^{+0.5}$ & $31.0_{-1.2}^{+0.8}$ & $70.3_{-1.2}^{+1.1}$ & $1.7_{-0.3}^{+0.3}$ & $11.8_{-1.6}^{+1.4}$  & $0.08_{-0.01}^{+0.01}$ &          ---           &            ---         \\[5pt]
IRAS 04302$+$2247 & I & $1.91_{-0.27}^{+0.12}$ & $325_{-13}^{+12}$ & $7.2_{-0.3}^{+0.3}$ & $86.8_{-0.7}^{+0.8}$ & $0.29_{-0.04}^{+0.05}$ & $6.4_{-0.6}^{+0.6}$ & $0.09_{-0.01}^{+0.01}$ & $1.21_{-0.15}^{+0.15}$ & $14.0_{-3.4}^{+3.4}Q$ \\[5pt]
\enddata
\tablenotetext{a}{Protostellar masses reported by Y. Aso et al. (2026, submitted).}
\tablenotetext{b}{$H_{d,0}$ is the dust scale height $H_{d}$ evaluated at $R = R_{0}$.}
\tablenotetext{c}{$H_{g,0}$ is the gas scale height $H_{g}$ evaluated at $R = R_{0}$. To calculate $H_{g,0}$, we assume that the gas also follows the dust temperature profile.}
\end{deluxetable*}

\section{Dust Settling-Triggered Substructure Formation During Class 0 to Class II Transition?}
\label{sec: Discussion}

\begin{figure}[t]
\centering
    \includegraphics[
      width=0.5\textwidth
    ]{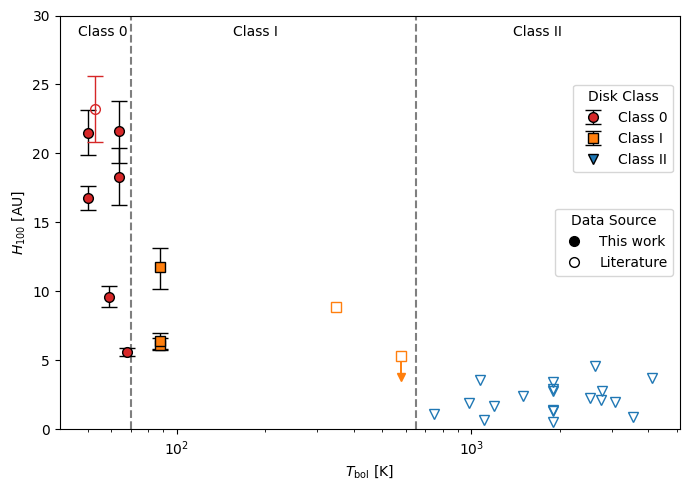}
    \caption{Dust scale height at $R = 100 \ \mathrm{AU}$, $H_{100}$, as a function of bolometric temperature $T_{\mathrm{bol}}$ for protostellar and protoplanetary disks spanning Class 0, Class I, and Class II evolutionary stages. Filled symbols represent sources modeled in this work, while open symbols denote values compiled from the literature; details of their derivation are provided in the text. Downward-pointing triangles denote upper limits. The dashed vertical lines mark the adopted boundaries between evolutionary classes based on $T_{\mathrm{bol}}$. The downward arrow associated with HL Tau, shown as the open square near the second vertical dashed line, indicates that its $H_{100}$ value is an upper limit. The overall decrease in $H_{100}$ as $T_{\mathrm{bol}}$ increases suggests progressive dust settling and disk flattening with evolutionary age.}
    \label{fig: H100_Disks}
\end{figure}

To place our results in an evolutionary context, we sort our modeled disks according to their bolometric temperatures, $T_{\mathrm{bol}}$, and compare them with the more evolved Class II disks in Figure~\ref{fig: H100_Disks}. Following \cite{Ohashi2023ApJ}, we adopt $T_{\mathrm{bol}} \sim 70~\mathrm{K}$ and $T_{\mathrm{bol}} \sim 650~\mathrm{K}$ as the boundaries between Class 0/I and Class I/II, respectively \citep{Evans2009ApJS,Ohashi2023ApJ}. The evolutionary comparison presented here builds on the qualitative picture suggested by \cite{Ohashi2023ApJ}, who noted that Class 0/I disks generally exhibit fewer prominent dust substructures than Class II disks and proposed that these substructures develop rapidly during the transition from the embedded to the Class II phase. Our results provide quantitative support for this evolutionary picture by showing that the dust layers in embedded disks remain geometrically thick, with little evidence for significant settling, and by suggesting that widespread dust settling and the emergence of prominent dust substructures occur over a similar evolutionary interval. For the Class 0/I sources modeled in this work, the bolometric temperatures are taken from \cite{Ohashi2023ApJ}. We compile those for HH~212 mms, HL Tau, and the Class II disks from the literature\footnote{For Class II sources without a measured bolometric temperature, we adopt a fiducial value of $T_{\mathrm{bol}} = 1900~\mathrm{K}$.}.

We begin by examining how the inferred dust scale height varies as a function of $T_{\mathrm{bol}}$. Figure \ref{fig: H100_Disks} shows the reference dust scale height $H_{100}$ at the reference radius of $R = 100 \ \mathrm{AU}$, with the data for all Class II disks (open triangles) taken from \cite{Villenave2025A&A}. We have included the Class 0 disk HH 212 mms in the figure (open circle), for which high-resolution ALMA observations and modeling revealed that the dust and gas are well mixed and therefore share the same vertical distribution \citep{Lin2021MNRAS}. Although the dust scale height was not explicitly modeled in that work, we estimated it by adopting $H_{d} = H_{g}$, where $H_{g}$ was calculated using the stellar mass adopted by \cite{Lin2021MNRAS} together with their best-fit outer disk radius and midplane temperature structure. We also include the Class I disk Oph IRS63 (the leftmost open square), whose dust scale height was inferred from radiative transfer modeling by S. Tobin et al. (2026, in preparation). For Oph IRS63, we adopted their ``smooth'' disk model in Band 6. We obtained the dust scale heights $H_{100}$ for the Class II disks from \cite{Villenave2025A&A}, which report the aspect ratio $H_{d}/R$ over a specified radial range. We included only sources for which the reported radial range contains or lies close to $R = 100 \ \mathrm{AU}$, allowing us to estimate $H_{100}$. One disk in their Class II sample is HL Tau, a borderline Class I object based on our adopted bolometric temperature classification; it is denoted by the open square near the second vertical dashed line separating the Class I and Class II populations. The uncertainty in $H_{d}$ for HH 212 mms was estimated by propagating the uncertainties in the model parameters reported by \cite{Lin2021MNRAS}. For literature values without reported uncertainties, we include only the central values.

Figure \ref{fig: H100_Disks} reveals a strong evolutionary trend: the inferred reference dust scale height $H_{100}$ decreases from the embedded Class 0 stage to Class I and becomes smallest in Class II disks\footnote{Encouragingly, an independent radiative transfer study of many of the same eDisk targets by Y. Choi et al. (in prep), carried out concurrently using a substantially different modeling framework, infers broadly similar dust scale heights for the disks common to the two studies. Their analysis simultaneously models ALMA Band 3 and Band 6 continuum emission and explicitly includes dust scattering, which our current models neglect. The broadly similar results increase confidence that the conclusion that embedded Class 0/I disks contain vertically extended dust with limited settling is robust.}. This likely reflects the combined effects of dust growth, settling, and declining dynamical stirring. In the youngest Class 0 systems, the dust has had the least time to grow and settle, and the disk is still being fed by envelope material that can continuously replenish small, well-coupled grains. These small grains are difficult to settle, especially in young systems where the central stellar masses may still be relatively low (Y. Aso et al. 2026, submitted), and the vertical gravitational pull toward the midplane is weaker. In addition, Class 0 disks are expected to have the highest accretion rates (N. Ohashi et al. 2026, in prep). They may therefore be more strongly stirred by turbulence, magnetic stresses, gravitational activity, or infall-driven motions, all of which can keep dust vertically extended. As the small-grain-bearing envelope dissipates and the disk accretion activity decreases, the grains can grow and settle more efficiently, leading to the thinner dust layers observed in Class II disks.

\begin{figure}[t]
\centering
    \includegraphics[
      width=0.5\textwidth
    ]{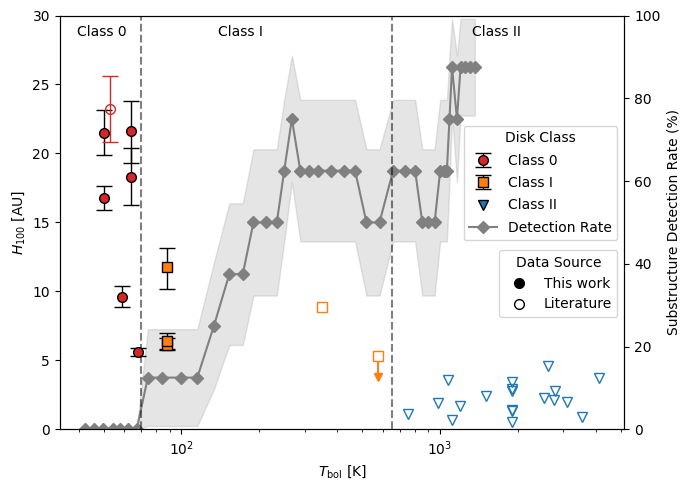}
    \caption{Same as Figure \ref{fig: H100_Disks}, now including the disk substructure detection rate from the CAMPOS survey analysis of \cite{Hsieh2025A&A} (gray diamonds and shaded region; right axis). The shaded region denotes the reported $1\sigma$ uncertainty from the counting statistics of the detection fraction. The increase in substructure occurrence toward higher $T_{\mathrm{bol}}$ coincides with lower measured dust scale heights, suggesting a connection between dust settling and the emergence of observable disk substructures.}
    \label{fig: H100_Sub_Disks}
\end{figure}

In addition to dust scale height, another dust property that shows a strong evolutionary trend is the fraction of disks with detected dust substructures, as illustrated in Figure \ref{fig: H100_Sub_Disks}. Specifically, the substructure detection rate increases strongly with the evolutionary stage, from essentially zero in the youngest Class 0 systems to nearly unity in many Class II disks. This trend was already suggested by the eDisk survey \citep{Ohashi2023ApJ}. More recently, \cite{Hsieh2025A&A} combined the CAMPOS sample with ALMA Band~6 data from the literature, including the eDisk sample, thereby providing stronger statistical support for the trend. Although the angular resolution of their sample is not homogeneous ($3$--$20~\mathrm{AU}$), they limited their analysis to sources for which the ratio of the apparent disk angular size to the angular resolution is greater than 4, allowing them to assess substructure detection more robustly. In addition, they limited their analysis to sources with inclination angles below 75 degrees to minimize geometric thickness effects in the dust. The identified substructures consist mainly of annular dust rings and gaps. Although the exact fraction of substructures may depend somewhat on how they are identified observationally, the overall evolutionary trend appears robust. The dust scale height measurements from Figure \ref{fig: H100_Disks} are overplotted on the substructure detection rates, revealing an intriguing anti-correlation between the dust scale height and substructure occurrence.

We should note that relatively few measurements of the dust scale height exist in the Class I range, making a direct comparison difficult. Nevertheless, the substructure detection rate appears to increase within Class I, specifically when $T_{\mathrm{bol}}$ is in the range $\sim 200$ to $400 \ \mathrm{K}$ \citep{Ohashi2023ApJ, Hsieh2025A&A, Narang2026AJ}. Additional measurements of the dust scale height for Class I disks could clarify whether an observational correlation exists, or whether dust settles before or after the emergence of dust substructures. In any case, there exists a strong contrast between the Class 0 and Class II extremes. Class 0 disks have large dust scale heights and a lack of substructure, whereas Class II disks have small dust scale heights and a high detection rate of substructures. This anti-correlation raises the intriguing possibility that dust settling may not merely accompany, but also facilitate, the emergence of annular substructures during disk evolution.

One possible mechanism linking dust settling with ring formation is the secular gravitational instability (SGI) of a thin dust layer embedded in gas \citep[e.g.,][]{Takahashi2014ApJ}. In this mechanism, gas drag dissipates epicyclic motions that would otherwise stabilize the dust layer, allowing dust self-gravity to amplify long-wavelength radial perturbations secularly. Global simulations by \cite{Tominaga2020ApJ} showed that SGI can naturally produce multiple dust rings qualitatively resembling many observed ALMA continuum rings. Dust settling may play a key role in this instability, since reducing the dust scale height increases the midplane dust density and strengthens the dust layer's self-gravity. In this scenario, the geometrically thick and dynamically active dust layers of Class 0 disks may remain largely stable against SGI. In contrast, the thinner, less vigorously stirred dust layers of Class II disks may become susceptible to the instability. The transition from Class 0 to Class II via the Class I phase may therefore mark the epoch when disks begin to settle and become dynamically quiet enough for SGI-driven ring formation to develop in earnest.

Dust settling may also facilitate substructure formation through streaming instability (SI). As solids settle toward the midplane, the local dust-to-gas ratio increases, strengthening the aerodynamic back-reaction of dust onto the gas. This back-reaction partially accelerates the gas toward Keplerian rotation, reducing the headwind experienced by drifting particles and slowing their inward radial migration. Regions with enhanced dust concentration therefore drift inward more slowly, allowing solids drifting in from neighboring regions at larger disk radii to accumulate and potentially produce long-lived radial pileups or ring-like dust concentrations. An even more exciting possibility is that dust-settling-triggered SI promotes rapid planetesimal formation and, eventually, planet formation. In this scenario, the observed increase in substructure occurrence from Class 0 to Class II disks could reflect, at least in part, the emergence of young planets that sculpt rings and gaps in the surrounding dust disk. Although the present data do not firmly establish such a causal sequence, the close correspondence between the onset of strong dust settling and the rapid rise in annular substructure detections suggests that dust settling may be a critical evolutionary step toward forming a structured, planet-forming disk.

\section{Conclusion}
\label{sec: Conclusion}

We investigated whether brightness asymmetries along the minor axis of highly inclined embedded protostellar disks can constrain their vertical dust distribution. Using axisymmetric radiative transfer calculations with RADMC-3D, we demonstrated that the observed asymmetry arises naturally from the combined effects of disk inclination, optical depth, and dust scale height. Although these parameters are partially degenerate, we show, using both simple geometric arguments and detailed radiative transfer modeling, that the observed minor-axis brightness asymmetry cannot be reproduced by a thin dust layer, independent of inclination and optical depth, and thus provides a robust indication that the dust is not completely settled. 

The near- and far-side orientations inferred from the dust continuum modeling of highly inclined eDisk Class 0/I disks are independently supported by outflow observations. Of the nine modeled disks, eight have useful outflow constraints, and in all eight cases, the modeled far side coincides with the side expected from the blueshifted outflow projection. This agreement independently validates the geometric interpretation underlying the use of minor-axis brightness asymmetry as a probe of dust vertical structure.

The best-fit models reproduce the observed continuum morphologies reasonably well and yield dust scale heights at $100 \ \mathrm{AU}$ ranging from $\sim 6$ to $\sim 22 \ \mathrm{AU}$. The inferred dust aspect ratios at the outer disk edge are typically $H_{d,0}/R_{0} \sim 0.05$--$0.16$. Most importantly, the dust scale heights are generally comparable to the gas scale heights, with $H_{d,0}/H_{g,0}$ close to unity for nearly all modeled sources. These results indicate that the millimeter-emitting dust in the embedded disks has experienced little, if any, settling toward the midplane.

When combined with literature results, we find evidence for a strong evolutionary trend: the inferred dust scale height decreases systematically from the youngest Class 0 stage to the most evolved Class II stage. This trend suggests that substantial dust settling occurs primarily during the Class I phase, when deeply embedded protostellar disks transition to revealed protoplanetary disks.

We further compared the inferred dust scale heights with the observed occurrence of disk substructures reported in recent surveys. Although the youngest Class 0 disks typically exhibit thick dust layers and few detected substructures, more evolved disks show both thinner dust layers and a substantially higher frequency of rings and gaps. Although the current data do not establish a causal connection, the correspondence suggests that dust settling may be a key evolutionary step toward forming structured planet-forming disks. Processes such as secular gravitational instability (SGI), streaming instability (SI), and planet formation itself may become more effective once solids are sufficiently concentrated toward the disk midplane.

Future high-resolution observations of larger samples of embedded disks, particularly during the Class I phase, will be essential to determine more precisely when significant dust settling begins and how rapidly it proceeds. Such observations will provide stronger constraints on the evolutionary pathway from vertically extended Class 0 disks to the thin dust layers commonly observed in Class II systems. They will help clarify whether dust settling and the emergence of disk substructures are indeed closely linked stages in the early evolution of planet-forming disks.

\section{Acknowledgement}
We thank the anonymous referee for the constructive comments that improved the manuscript. The authors thank Cheng-Han Hsieh for providing substructure detection rates across disk evolutionary stages and Chun-Yen Hsu for helpful discussions. We acknowledge access to computational resources through the RIVANNA supercomputer facility at the University of Virginia and NASA High-Performance Computing. S.K.S. and Z.-Y.L. are supported in part by NASA 80NSSC20K0533, NSF AST-2307199, JWST-GO-02104.002-A, JWST-GO-08872.003-A, and the Virginia Institute of Theoretical Astronomy (VITA). L.W.L. acknowledges support from NSF AST-2108794.  N.O. acknowledges support from the National Science and Technology Council (NSTC) in Taiwan through the grants NSTC 114-2112-M-001-019 and NSTC 115-2112-M-001-008, and also from the Academia Sinica Investigator Project grant (AS-IV-114-M02). I.H. acknowledges funding from the European Research Council (ERC) under the European Union’s Horizon 2020 research and innovation program (grant agreement No. 101098309 - PEBBLES).

\appendix
\section{Modeling Results for the Full Sample}
\label{sec: Full_Sample}
For completeness, we present the observed images, best-fit models, and residual maps for all sources in our sample not included in the main text. The figures follow the same format as Figure \ref{fig: 2D_RADMC3D_GSS30IRS3} for GSS30 IRS3.
\begin{figure*}[t]
\centering
    \includegraphics[
      width=1.0\textwidth
    ]{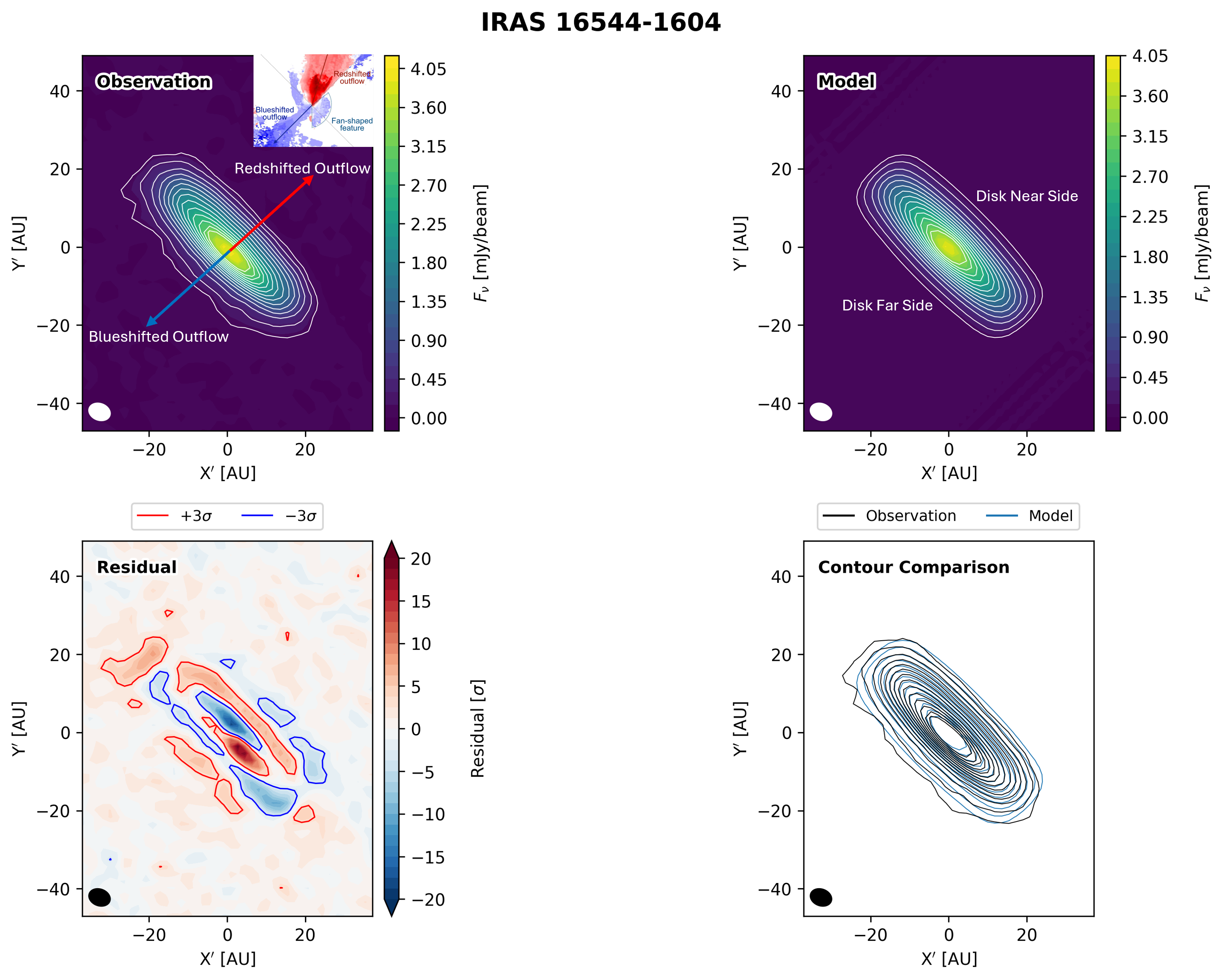}
    \caption{Comparison between the observed and modeled continuum emission for IRAS 16544$-$1604 (also known as CB 68). The panels follow the same format as Figure \ref{fig: 2D_RADMC3D_GSS30IRS3}: observed continuum image, best-fit RADMC-3D model, residual map, and observed/model contour comparison. In the observation panel, the inset displays the CO molecular outflow from \cite{Kido2023ApJ}, with the blue and red arrows indicating the projected directions of the blueshifted and redshifted outflow lobes, respectively. The modeled near and far sides of the disk are labeled in the model panel, and their orientation is consistent with the independent expectation from the observed outflow geometry. In the residual map, red and blue contours mark the $+3\sigma$ and $-3\sigma$ residual levels, respectively. In the contour-comparison panel, black contours show the observed emission and blue contours show the model emission. The same contour levels are used for the observed and modeled images: the lowest contour is set at $5\sigma$, while the remaining contours are linearly spaced up to $80\%$ of the observed peak intensity, corresponding approximately to $[11, 23, \ldots, 130, 142, 154]\sigma$.}
    \label{fig: 2D_RADMC3D_CB68}
\end{figure*}
\begin{figure*}[t]
\centering
    \includegraphics[
      width=1.0\textwidth
    ]{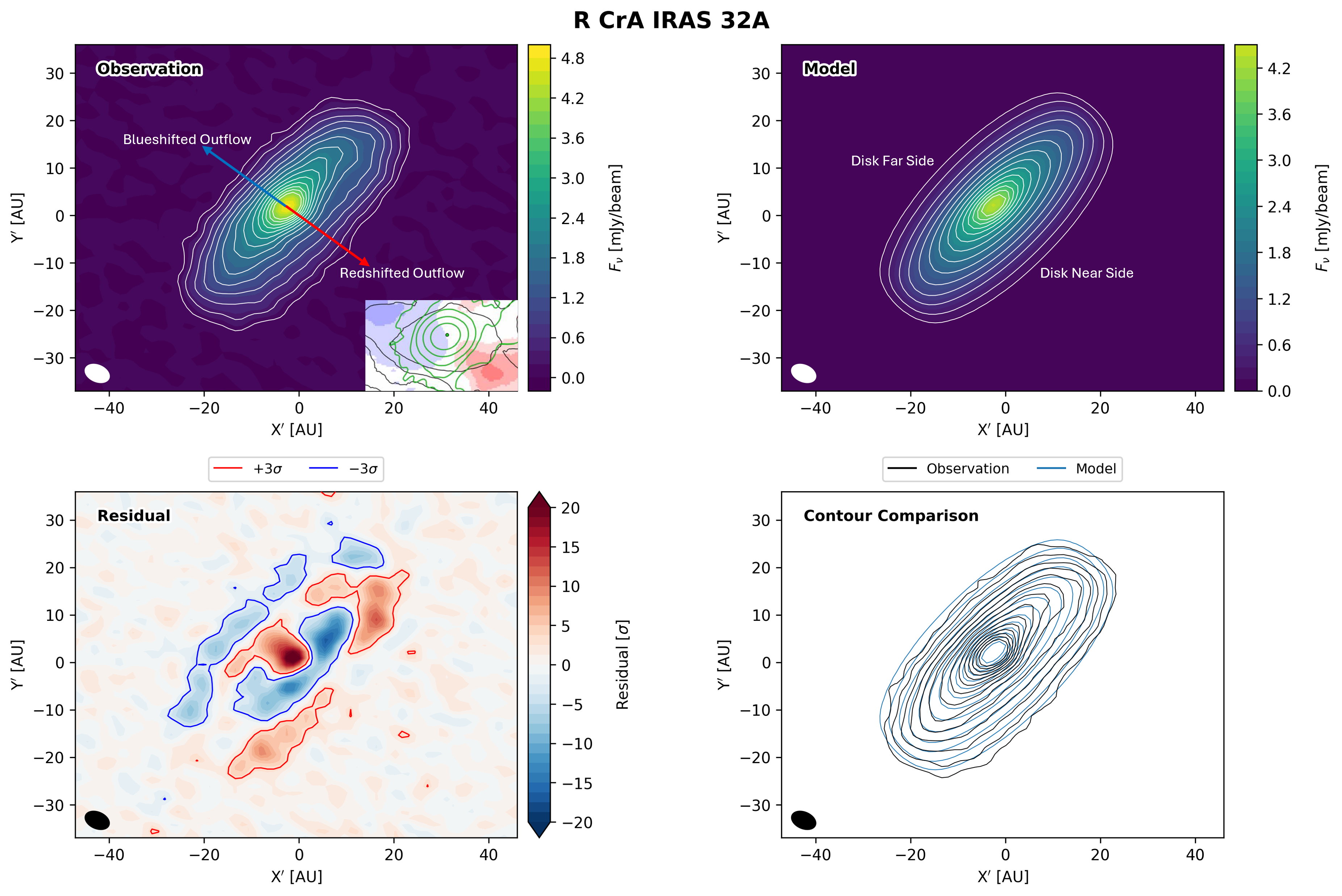}
    \caption{Comparison between the observed and modeled continuum emission for R CrA IRAS 32A. The panels follow the same format as Figure \ref{fig: 2D_RADMC3D_GSS30IRS3}: observed continuum image, best-fit RADMC-3D model, residual map, and observed/model contour comparison. In the observation panel, the inset displays the CO molecular outflow from \cite{Encalada2024ApJ}, with the blue and red arrows indicating the projected directions of the blueshifted and redshifted outflow lobes, respectively. The modeled near and far sides of the disk are labeled in the model panel, and their orientation is consistent with the independent expectation from the observed outflow geometry. In the residual map, red and blue contours mark the $+3\sigma$ and $-3\sigma$ residual levels, respectively. In the contour-comparison panel, black contours show the observed emission and blue contours show the model emission. The same contour levels are used for the observed and modeled images: the lowest contour is set at $5\sigma$, while the remaining contours are linearly spaced up to $80\%$ of the observed peak intensity, corresponding approximately to $[11, 23, \ldots, 128, 140, 152]\sigma$.}
    \label{fig: 2D_RADMC3D_IRAS32A}
\end{figure*}
\begin{figure*}[t]
\centering
    \includegraphics[
      width=1.0\textwidth
    ]{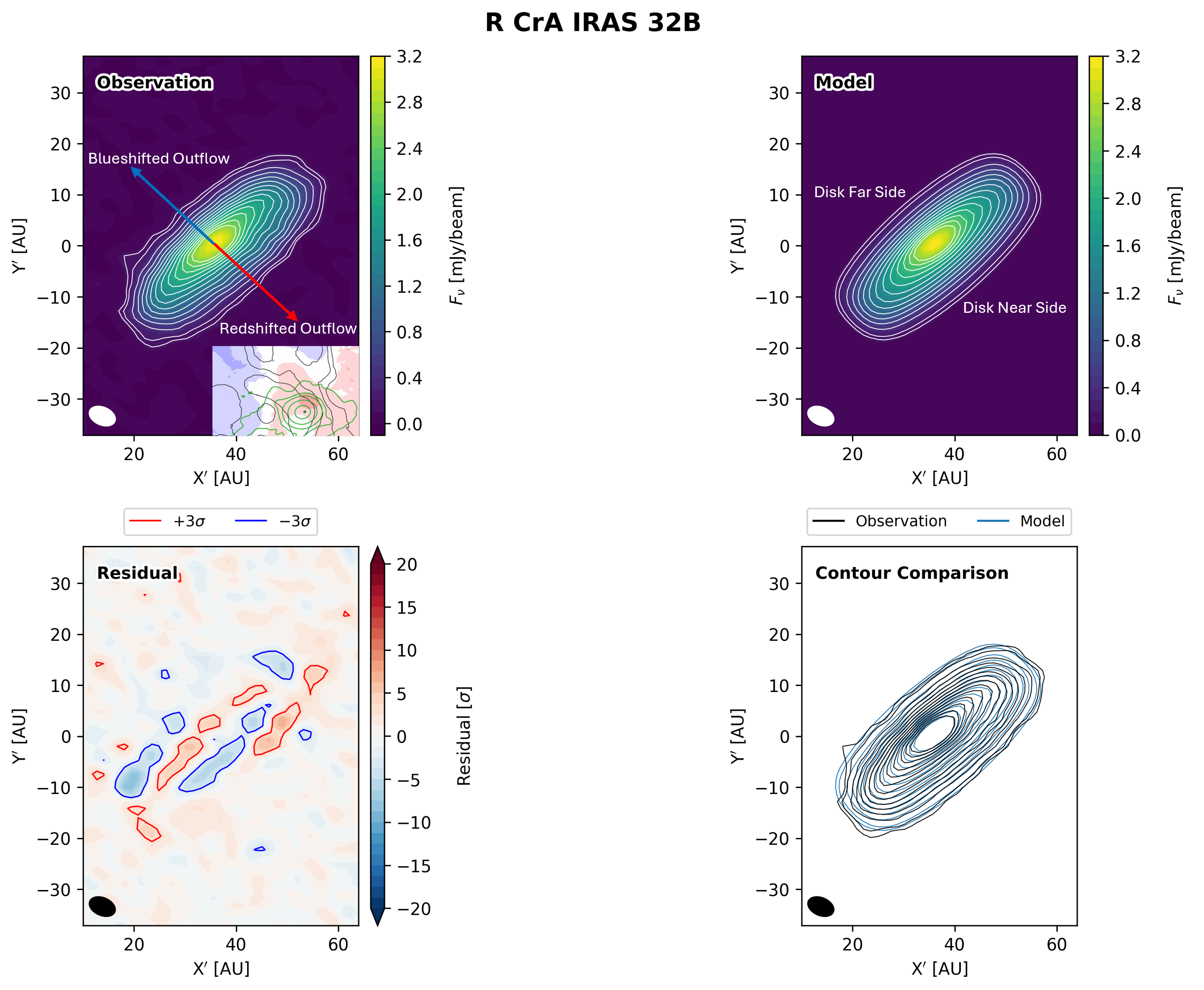}
    \caption{Comparison between the observed and modeled continuum emission for R CrA IRAS 32B. The panels follow the same format as Figure \ref{fig: 2D_RADMC3D_GSS30IRS3}: observed continuum image, best-fit RADMC-3D model, residual map, and observed/model contour comparison. In the observation panel, the inset displays the CO molecular outflow from \cite{Encalada2024ApJ}, with the blue and red arrows indicating the projected directions of the blueshifted and redshifted outflow lobes, respectively. The modeled near and far sides of the disk are labeled in the model panel, and their orientation is consistent with the independent expectation from the observed outflow geometry. In the residual map, red and blue contours mark the $+3\sigma$ and $-3\sigma$ residual levels, respectively. In the contour-comparison panel, black contours show the observed emission and blue contours show the model emission. The same contour levels are used for the observed and modeled images: the lowest contour is set at $5\sigma$, while the remaining contours are linearly spaced up to $80\%$ of the observed peak intensity, corresponding approximately to $[7, 15, \ldots, 84, 91, 99]\sigma$.}
    \label{fig: 2D_RADMC3D_IRAS32B}
\end{figure*}
\begin{figure*}[t]
\centering
    \includegraphics[
      width=1.0\textwidth
    ]{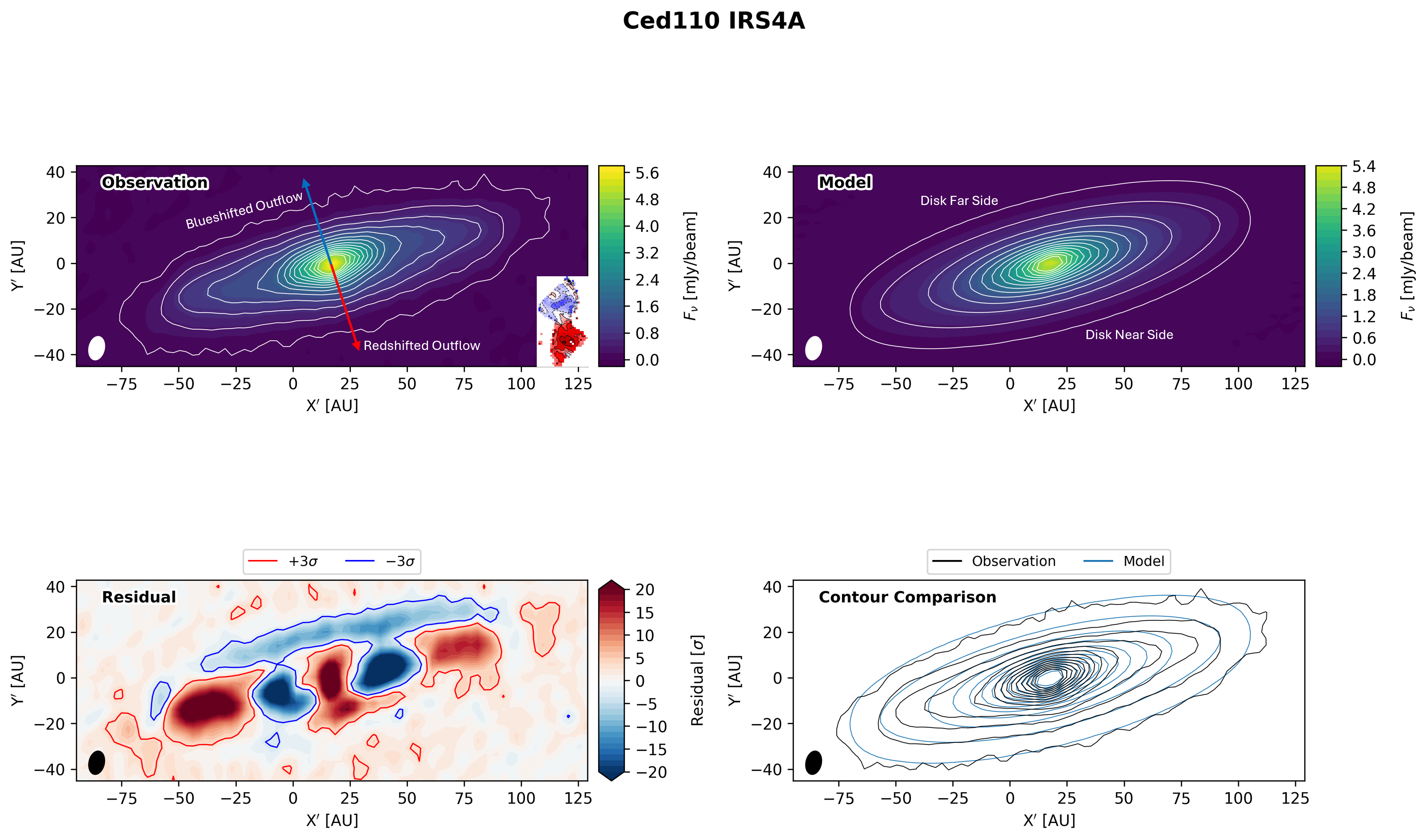}
    \caption{Comparison between the observed and modeled continuum emission for Ced110 IRS4A. The panels follow the same format as Figure \ref{fig: 2D_RADMC3D_GSS30IRS3}: observed continuum image, best-fit RADMC-3D model, residual map, and observed/model contour comparison. In the observation panel, the inset displays the JWST-detected [Fe II] atomic jet from \cite{Narang2025AJ}, with the blue and red arrows indicating the projected directions of the blueshifted and redshifted outflow lobes, respectively. The modeled near and far sides of the disk are labeled in the model panel, and their orientation is consistent with the independent expectation from the observed outflow geometry. In the residual map, red and blue contours mark the $+3\sigma$ and $-3\sigma$ residual levels, respectively. In the contour-comparison panel, black contours show the observed emission and blue contours show the model emission. The same contour levels are used for the observed and modeled images: the lowest contour is set at $5\sigma$, while the remaining contours are linearly spaced up to $80\%$ of the observed peak intensity, corresponding approximately to $[23, 47, \ldots, 258, 282, 305]\sigma$.}
    \label{fig: 2D_RADMC3D_IRS4A}
\end{figure*}
\begin{figure*}[t]
\centering
    \includegraphics[
      width=1.0\textwidth
    ]{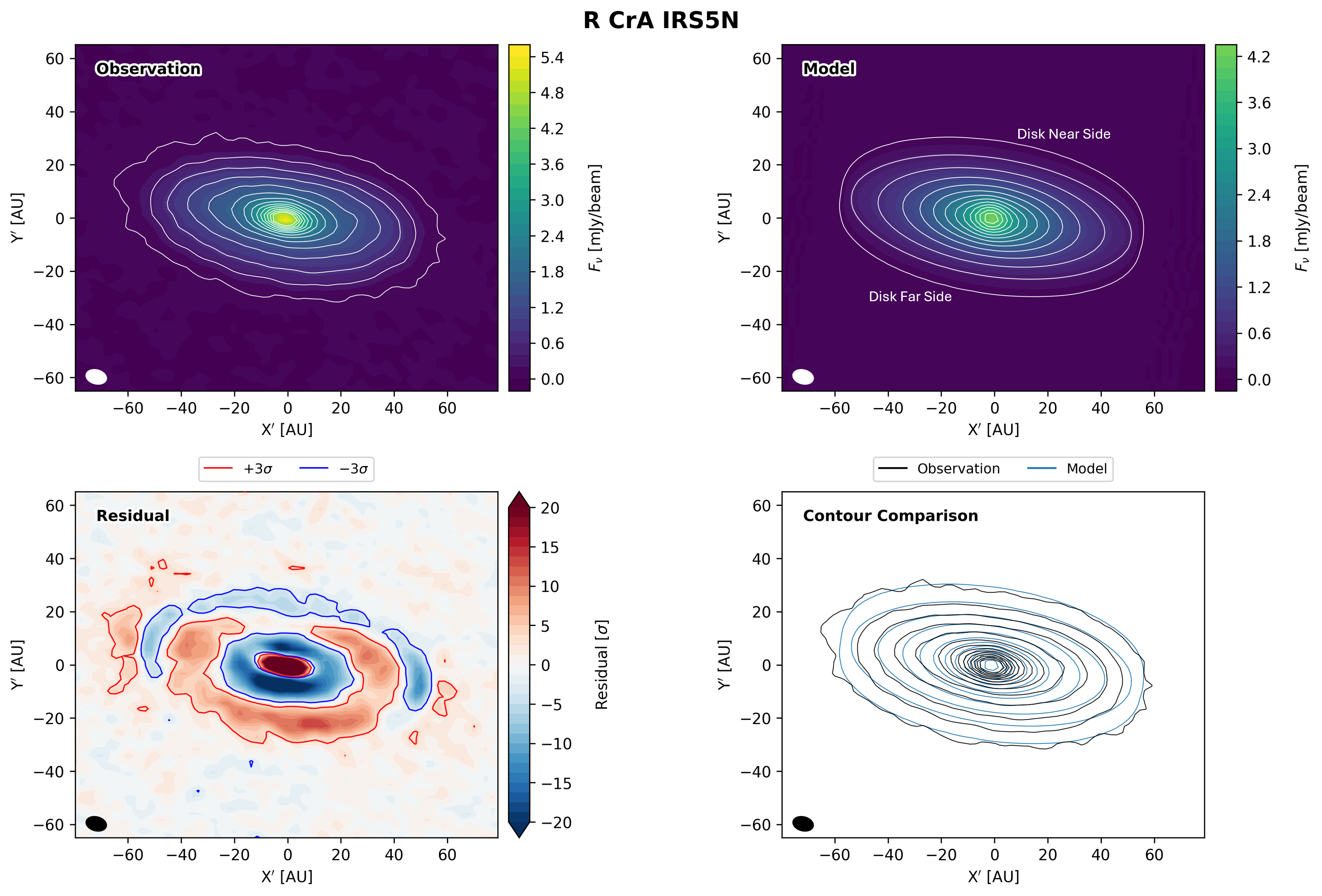}
    \caption{Comparison between the observed and modeled continuum emission for R CrA IRS5N. The panels follow the same format as Figure \ref{fig: 2D_RADMC3D_GSS30IRS3}: observed continuum image, best-fit RADMC-3D model, residual map, and observed/model contour comparison. No securely identified bipolar outflow approximately aligned with the projected disk minor axis is available for this source; therefore, the modeled near- and far-side orientation labeled in the model panel cannot be independently assessed from the outflow geometry. In the residual map, red and blue contours mark the $+3\sigma$ and $-3\sigma$ residual levels, respectively. In the contour-comparison panel, black contours show the observed emission and blue contours show the model emission. The same contour levels are used for the observed and modeled images: the lowest contour is set at $5\sigma$, while the remaining contours are linearly spaced up to $80\%$ of the observed peak intensity, corresponding approximately to $[24, 49, \ldots, 270, 295, 320]\sigma$.}
    \label{fig: 2D_RADMC3D_IRS5N}
\end{figure*}
\begin{figure*}[t]
\centering
    \includegraphics[
      width=1.0\textwidth
    ]{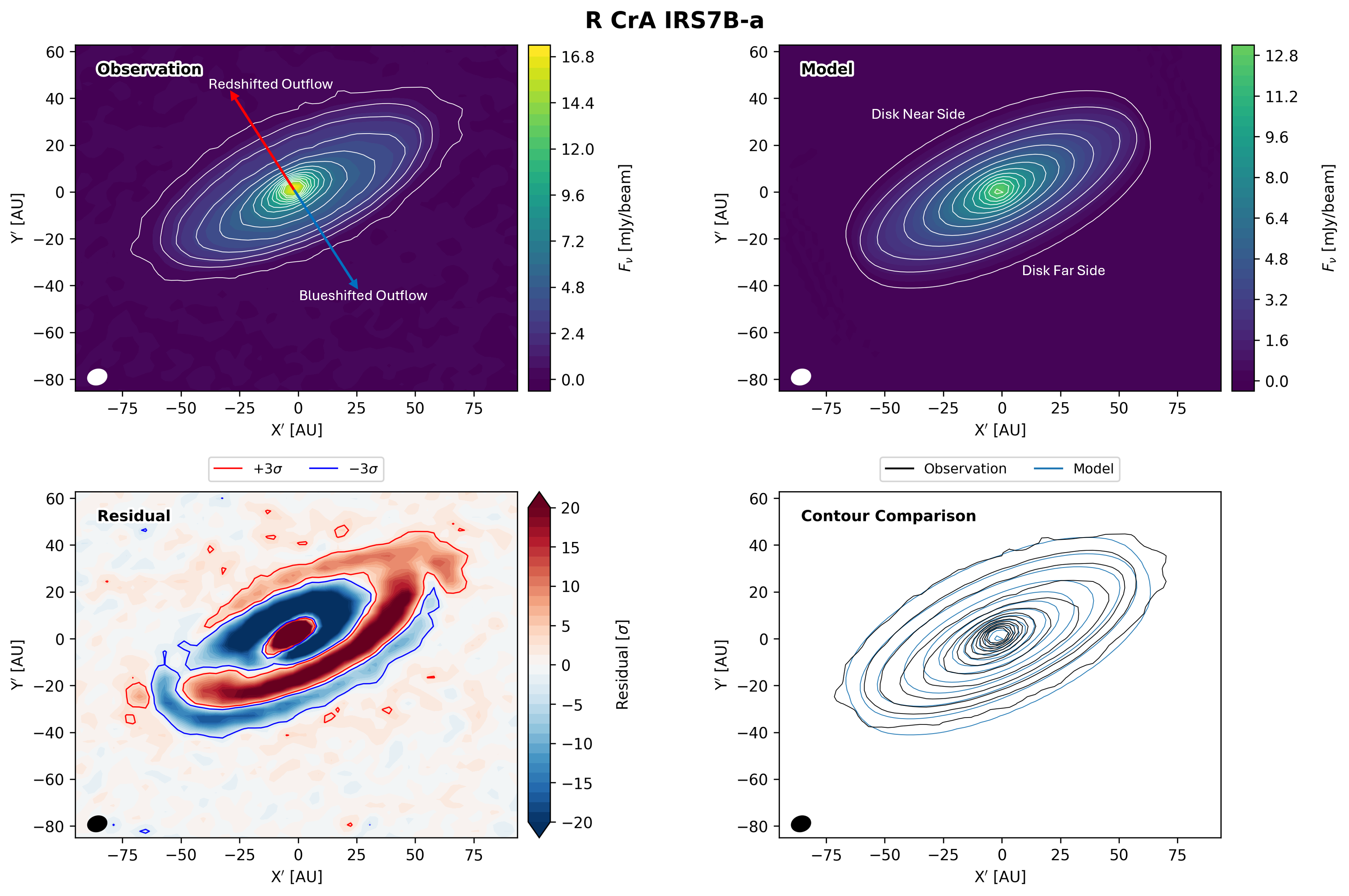}
    \caption{Comparison between the observed and modeled continuum emission for R CrA IRS7B$-$a. The panels follow the same format as Figure \ref{fig: 2D_RADMC3D_GSS30IRS3}: observed continuum image, best-fit RADMC-3D model, residual map, and observed/model contour comparison. In the observation panel, the blue arrow indicates the projected direction of the blueshifted jet/outflow toward the southwest, as inferred indirectly from the blueshifted molecular arc, shock emission, and jet-driven cavity reported by \cite{Sabatini2024A&A}. The red arrow indicates the ``expected'' opposite lobe direction. The modeled near and far sides of the disk are labeled in the model panel, and their orientation is consistent with the independent expectation from the observed outflow geometry. In the residual map, red and blue contours mark the $+3\sigma$ and $-3\sigma$ residual levels, respectively. In the contour-comparison panel, black contours show the observed emission, and blue contours show the model emission. The same contour levels are used for the observed and modeled images: the lowest contour is set at $5\sigma$, while the remaining contours are linearly spaced up to $80\%$ of the observed peak intensity, corresponding approximately to $[33, 67, \ldots, 373, 407, 441]\sigma$. We note that \cite{Sabatini2024A&A} reported the outflow from R CrA IRS7B as an unresolved system, so the arrows indicate the large-scale IRS7B outflow geometry rather than a component-specific outflow.}
    \label{fig: 2D_RADMC3D_IRS7B-A}
\end{figure*}
\begin{figure*}[t]
\centering
    \includegraphics[
      width=1.0\textwidth
    ]{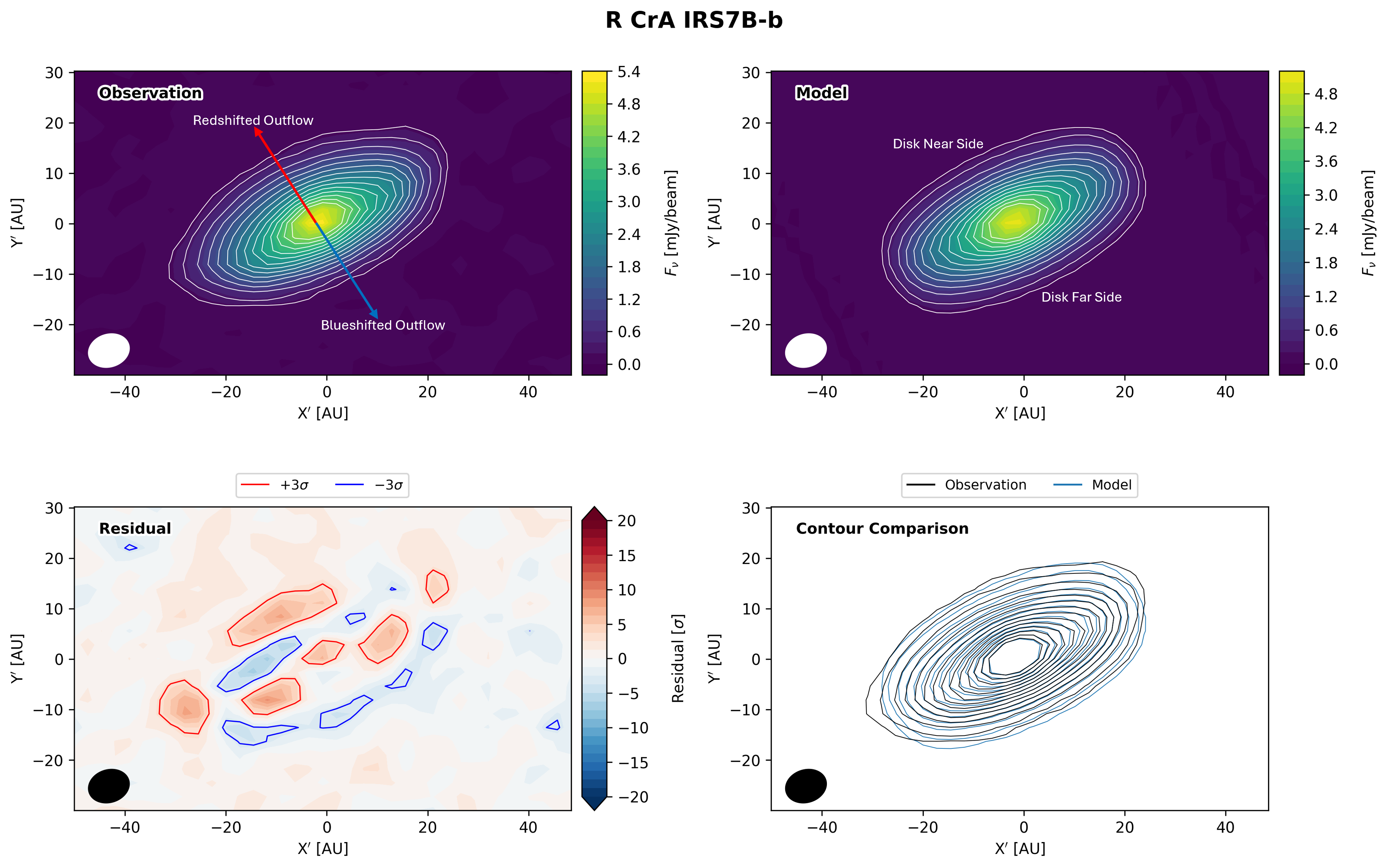}
    \caption{Comparison between the observed and modeled continuum emission for R CrA IRS7B$-$b. The panels follow the same format as Figure \ref{fig: 2D_RADMC3D_GSS30IRS3}: observed continuum image, best-fit RADMC-3D model, residual map, and observed/model contour comparison. In the observation panel, the blue arrow indicates the projected direction of the blueshifted jet/outflow toward the southwest, as inferred from the blueshifted molecular arc, shock emission, and jet-driven cavity reported by \cite{Sabatini2024A&A}. The red arrow indicates the ``expected'' opposite lobe direction. The modeled near and far sides of the disk are labeled in the model panel, and their orientation is consistent with the independent expectation from the observed outflow geometry. In the residual map, red and blue contours mark the $+3\sigma$ and $-3\sigma$ residual levels, respectively. In the contour-comparison panel, black contours show the observed emission and blue contours show the model emission. The same contour levels are used for the observed and modeled images: the lowest contour is set at $5\sigma$, while the remaining contours are linearly spaced up to $80\%$ of the observed peak intensity, corresponding approximately to $[10, 21, \ldots, 115, 126, 136]\sigma$. We note that \cite{Sabatini2024A&A} reported the outflow from R CrA IRS7B as an unresolved system, so the arrows indicate the large-scale IRS7B outflow geometry rather than a component-specific outflow.}
    \label{fig: 2D_RADMC3D_IRS7B-B}
\end{figure*}
\begin{figure*}[t]
\centering
    \includegraphics[
      width=0.7\textwidth
    ]{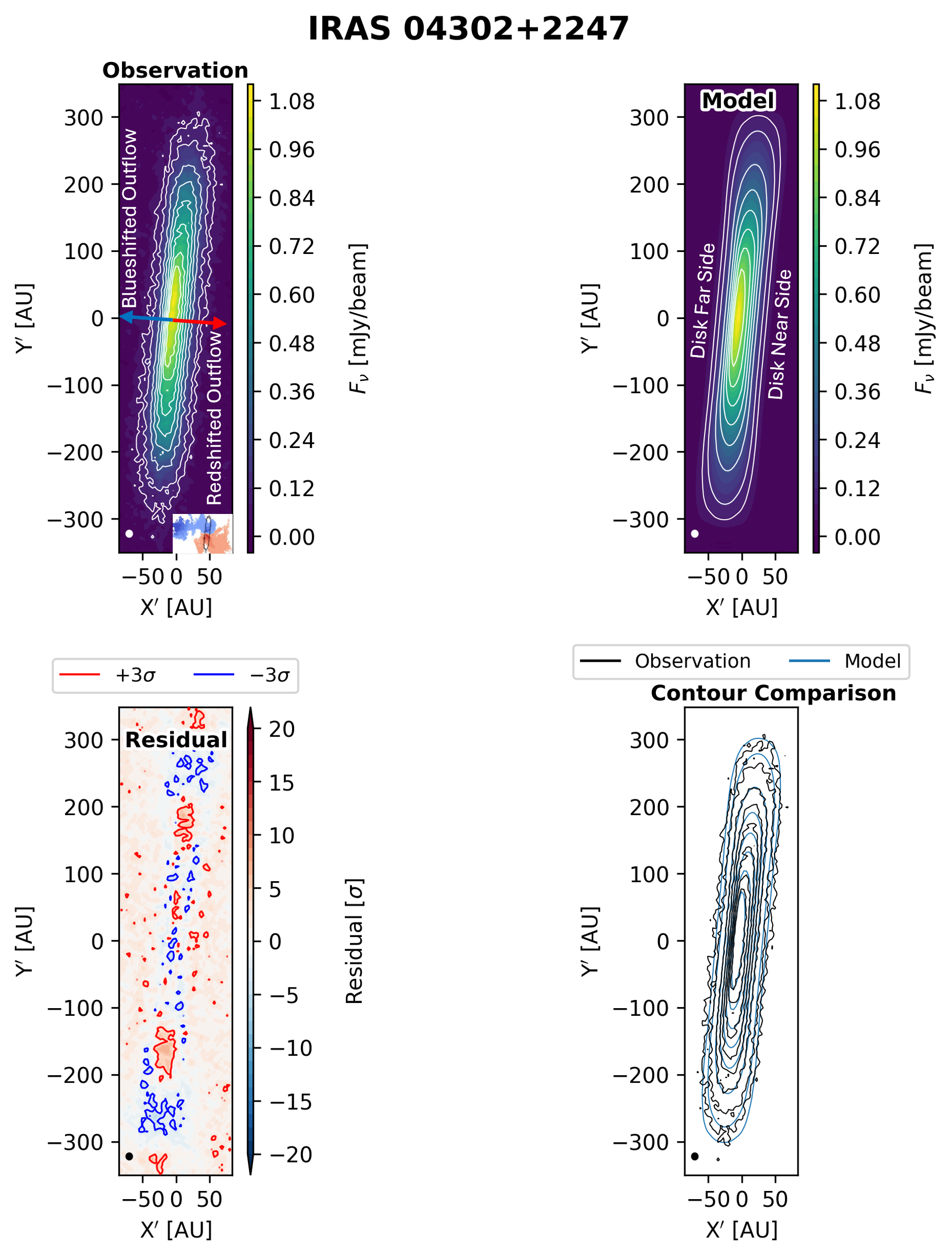}
    \caption{Comparison between the observed and modeled continuum emission for IRAS 04302$+$2247. The panels follow the same format as Figure \ref{fig: 2D_RADMC3D_GSS30IRS3}: observed continuum image, best-fit RADMC-3D model, residual map, and observed/model contour comparison. In the observation panel, the inset displays the CO molecular outflow from \cite{Lin2023ApJ}, with the blue and red arrows indicating the projected directions of the blueshifted and redshifted outflow lobes, respectively. The modeled near and far sides of the disk are labeled in the model panel, and their orientation is consistent with the independent expectation from the observed outflow geometry. In the residual map, red and blue contours mark the $+3\sigma$ and $-3\sigma$ residual levels, respectively. In the contour-comparison panel, black contours show the observed emission and blue contours show the model emission. The same contour levels are used for the observed and modeled images: the lowest contour is set at $5\sigma$, while the remaining contours are linearly spaced up to $80\%$ of the observed peak intensity, corresponding approximately to $[8, 17, \ldots, 43, 51, 60]\sigma$.}
    \label{fig: 2D_RADMC3D_IRAS04302}
\end{figure*}

\clearpage
\bibliographystyle{aasjournalv7}
\bibliography{main}
\end{document}